\documentclass[twocolumn,superscriptaddress,float,prl]{revtex4-1}
\usepackage{graphicx,amsfonts,amssymb,amsmath,hyperref,hypcap,enumerate}
\usepackage{color}
\usepackage{cleveref}

\newcommand{\qq}{\boldsymbol{q}}
\newcommand{\QQ}{\boldsymbol{Q}}

\newcommand{\rr}{\boldsymbol{r}}

\newcommand{\kk}{\boldsymbol{k}}

\newcommand{\bb}{\boldsymbol{b}}

\newcommand{\mm}{\boldsymbol{m}}
\newcommand{\PP}{\boldsymbol{\pi}}

\begin{document}
\title{Collective Excitations of Quantum Anomalous Hall Ferromagnets \\
	in Twisted Bilayer Graphene}

\author{Fengcheng Wu}
%\email{wufcheng@umd.edu}
\affiliation{Condensed Matter Theory Center and Joint Quantum Institute, Department of Physics, University of Maryland, College Park, Maryland 20742, USA}
\author{Sankar Das Sarma}
\affiliation{Condensed Matter Theory Center and Joint Quantum Institute, Department of Physics, University of Maryland, College Park, Maryland 20742, USA}

%\date{\today}

\begin{abstract}
	We present a microscopic theory for collective excitations of quantum anomalous Hall ferromagnets (QAHF) in twisted bilayer graphene. We calculate the spin magnon and valley magnon spectra by solving Bethe-Salpeter equations, and verify the stability of QAHF. We extract the spin stiffness from the gapless spin wave dispersion, and estimate the energy cost of a skyrmion-antiskyrmion pair, which is found to be comparable in energy with the Hartree-Fock gap. The valley wave mode is gapped, implying that the valley polarized state is more favorable compared to the valley coherent state. Using a nonlinear sigma model, we estimate the valley ordering temperature, which is considerably reduced from the mean-field transition temperature due to thermal excitations of valley waves.
\end{abstract}

\maketitle

%\section{introduction}
{\it Introduction.---}
Twisted bilayer graphene (TBG) near the magic angle hosts a plethora of phenomena, e.g., superconductivity \cite{Cao2018Super}, correlated insulators\cite{Cao2018Magnetic}, nematicity \cite{Kerelsky2019,choi2019imaging}, large linear-in-temperature resistivity\cite{MIT2018_rho,Columbia2018_rho}, quantum anomalous Hall effect (QAHE)\cite{sharpe2019emergent,serlin2019intrinsic}, etc. Due to this richness, TBG and related moir\'e systems are currently under intense experimental \cite{sharpe2019emergent,serlin2019intrinsic,chen2019tunable,Cao2018Super,Cao2018Magnetic,Kerelsky2019,choi2019imaging,MIT2018_rho,Columbia2018_rho,Dean2018tuning,codecido2019correlated,lu2019superconductors,tomarken2019electronic,Xie2019,Jiang2019,shen2019observation,liu2019spin,cao2019electric} and theoretical  \cite{Balents2018,Senthil2018,Koshino2018, Kang2018, Liu2018chiral, Dodaro12018,Isobe2018,You2018, Tang2019, rademaker2018charge, guinea2018electrostatic, gonzalez2018kohn, Lin2018,Lado2018,Vishwanath2018origin, Ahn2018failure,Bernevig2018Topology, hejazi2018multiple,  sherkunov2018novel,kang2018strong,Seo2019,lin2019chiral,Heikkila2018, Lian2018twisted, choi2018electron,Wu2018phonon, wu2019phonon,wu2018topological,wu2019identification,wu2019Ferro, Zhang2019,JungTopology,WuTITMD,xie2018nature,bultinck2019anomalous,zhang2019twisted,LiuMulti,lee2019theory,Xu2019Ferro,hazra2018upper,xie2019topology,julku2019superfluid,hu2019geometric,liu2019anomalous} study. For QAHE, which is the focus of this work, moir\'e bilayers emerge as a new and clean system \cite{serlin2019intrinsic,chen2019tunable} to realize  Chern insulators at elevated temperatures compared with the magnetic topological insulators \cite{Chang167}.

Moir\'e superlattices in van der Waals bilayers not only generate nearly flat bands, but also often endow the bands with nontrivial topology. In moir\'e systems with valley contrast Chern numbers, the enhanced electron Coulomb repulsion effect due to band flattening can spontaneously break the valley degeneracy and therefore, lead to valley polarized states with QAHE \cite{Zhang2019,JungTopology,WuTITMD,xie2018nature,bultinck2019anomalous,zhang2019twisted,LiuMulti}; we term such bulk insulating states as quantum anomalous Hall ferromagnets (QAHF), in analogy with the well known quantum Hall ferromagnets (QHF) \cite{Moon1995,Yang2006}. In pristine TBG, $\hat{C}_{2z}$ symmetry (a two-fold rotation around the out-of-plane axis) combined with time-reversal symmetry forbids Berry curvature. However, this $\hat{C}_{2z}$ symmetry can be explicitly broken when TBG is aligned to the hexagonal boron nitride (hBN) substrate, generating a nonzero valley Chern number.  It is in this extrinsic TBG aligned with hBN where the anomalous Hall effect (AHE) \cite{sharpe2019emergent} and later its quantized version (QAHE) \cite{serlin2019intrinsic} have been observed at the filling factor $\nu=3$. Here we define $\nu$  as $n/n_0$, where $n$ is the electron density, and $n_0$ the density for one electron per moir\'e  unit cell.

In this paper, we theoretically study the collective excitations in the TBG QAHF, in order to examine the QAHF stability, and to determine the low-energy excitations that control the transport gap and that limit the ferromagnetic transition temperature. The $\nu=3$ QAHF in extrinsic TBG has two distinct collective excitations, i.e., spin magnons and valley magnons, which involve particle-hole transitions with respectively, a single spin flip and a single valley flip. We calculate the energy spectra separately for the two types of magnons by solving their Bethe-Salpeter equations. The calculated excitation spectra indicate that the TBG QAHF is generally robust against small particle-hole fluctuations when the bulk Hartree-Fock gap ($\Delta_{\text{HF}}$) is finite. The spin magnon spectrum has a gapless spin wave mode, which is the Goldstone mode due to the spontaneous breaking of the spin SU(2) symmetry in the $\nu=3$ QAHF. We extract spin stiffness from the long-wavelength spin wave dispersion, and estimate the skyrmion energy. We find that the energy $\Delta_{\text{pair}}$ for a pair of skyrmion and antiskyrmion in the TBG QAHF is comparable in energy with $\Delta_{\text{HF}}$, and either $\Delta_{\text{pair}}$ or $\Delta_{\text{HF}}$ can be the lowest charged excitation gap depending on details of the system.
%Because the emergent QAHF does not break any continuous symmetry in the valley pseudospin space, the valley magnon spectrum is fully gapped and positive in energy, which implies that the valley polarized state is energetically favorable compared to the valley coherent state.

In a two-dimensional system such as TBG with spin SU(2) symmetry, the spin ordering temperature vanishes according to the Mermin-Wagner theorem. However, QAHE in TBG can arise purely from an orbital effect, e.g., valley polarization. An important distinction between spin and valley is that there is only a valley U(1) symmetry in TBG in contrast to the spin SU(2)  symmetry. The $\nu=3$ QAHF preserves the valley U(1) symmetry, but breaks the discrete time-reversal symmetry, which allows a finite valley ordering temperature $T_V$. We estimate $T_V$ based on the fully gapped valley magnon spectrum, and find that $T_V$ is reduced from the mean-field transition temperature due to thermal excitations of valley waves, which provides an explanation for the experimentally observed hierarchy that the transport energy gap of the TBG QAHF is larger than the corresponding Curie temperature \cite{serlin2019intrinsic}.

\begin{figure}[t]
	\includegraphics[width=1\columnwidth]{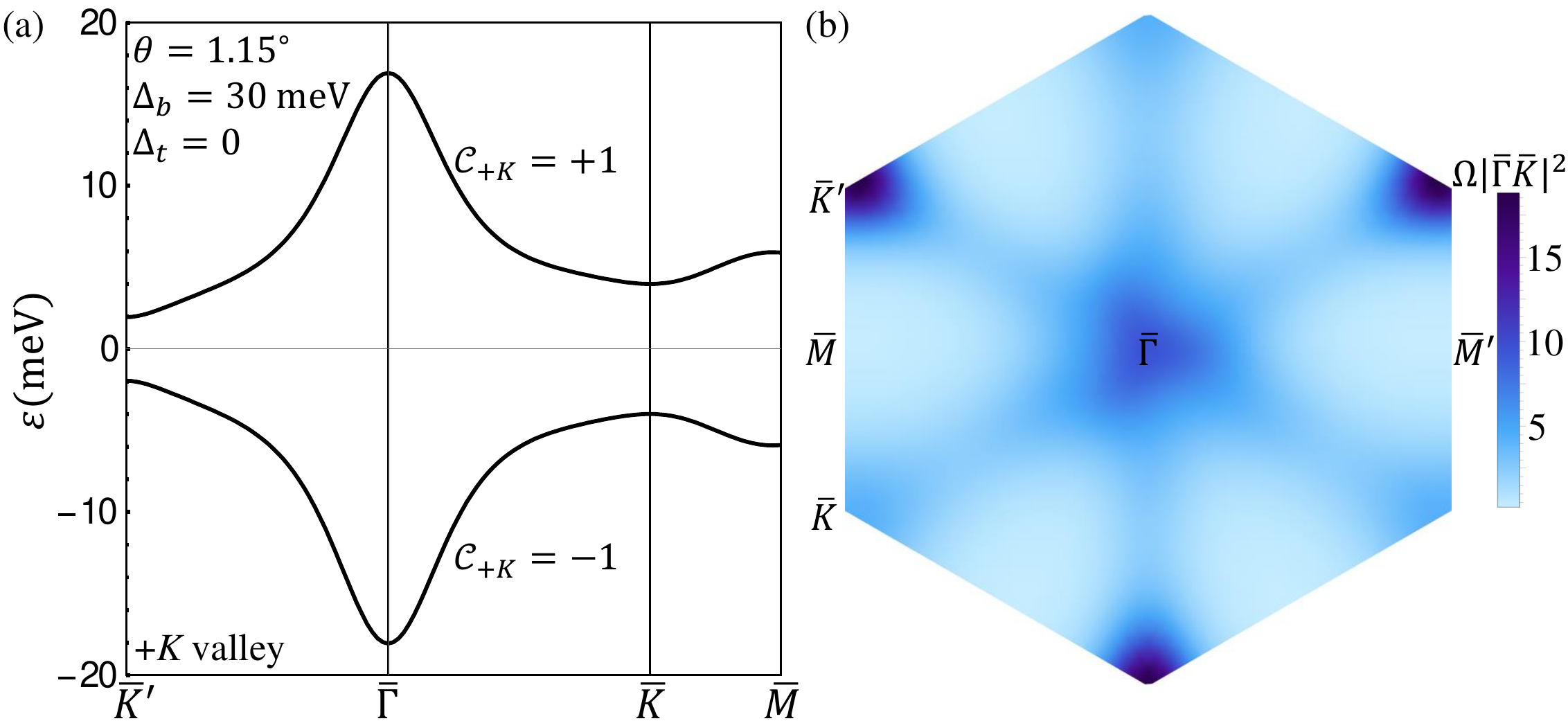}
	\caption{(a)The $+K$-valley moir\'e band structure with $\theta=1.15^{\circ}$ and $(\Delta_b, \Delta_t)=(30, 0)$ meV . (b) Berry curvature $\Omega$ of the first moir\'e conduction band in (a). We use a definition of $\Omega$ such that an occupied band with a Chern number $\mathcal{C}$ contributes $+\mathcal{C} e^2/h$ to the Hall conductivity $\sigma_{xy}$.}
	\label{Fig:band}
\end{figure}

%\section{Moir\'e Bands}
%\label{section:moire}

{\it Ferromagnetism.---} We calculate the moir\'e band structure of TBG using the continuum Hamiltonian \cite{Bistritzer2011}, with details given in the Supplemental Material (SM) \cite{SM}. We use parameter $\Delta_b$ ($\Delta_t$) to describe the sublattice potential difference in the bottom (top) graphene layer, and take $(\Delta_b, \Delta_t)=(30, 0)$ meV  \cite{Hunt2013} in order to simulate the experimental situation \cite{sharpe2019emergent,serlin2019intrinsic} where TBG is in close alignment to one of the two (either top or bottom) encapsulating hBN layers.  The corresponding moir\'e band structure in $+K$-valley at twist angle $\theta=1.15^{\circ}$ is shown in Fig.~\ref{Fig:band}, where the first moir\'e conduction and valence bands are separated by an energy gap about 4 meV (opened up by $\Delta_b$), and respectively carry a Chern number  $\mathcal{C}$ of $+1$ and $-1$. Because of time reversal symmetry, the first moir\'e conduction (valence) band in $-K$ valley has a $\mathcal{C}$ value of $-1$($+1$). 

We study a minimal interacting model by retaining only the first moir\'e conduction band states, assuming that all valence band states are filled. The projected Hamiltonian $H$ has the single-particle term $H_0$ and the interacting term $H_1$,
\begin{equation}
\begin{aligned}
&H_0= \sum_{\kk, \tau, s} \varepsilon_{\kk,\tau} c^{\dagger}_{\kk,\tau, s} c_{\kk,\tau, s}\\
&H_1=\frac{1}{2A}\sum V_{\kk_1 \kk_2 \kk_3 \kk_4}^{(\tau \tau')}
c^{\dagger}_{\kk_1,\tau, s} c^{\dagger}_{\kk_2,\tau', s'} c_{\kk_3,\tau', s'} c_{\kk_4,\tau, s}, \\
&V_{\kk_1 \kk_2 \kk_3 \kk_4}^{(\tau \tau')} = \sum_{\qq} V(\qq) O_{\kk_1 \kk_4}^{(\tau)} (\qq) O_{\kk_2 \kk_3}^{(\tau')}(-\qq), \\
&O_{\kk \kk'}^{(\tau)} (\qq) =  \int d\rr e^{i \qq\cdot \rr} \Phi_{\kk,\tau}^{*}(\rr) \Phi_{\kk',\tau}(\rr),
\end{aligned}
\end{equation}
where $c^{\dagger}_{\kk,\tau, s}$, $\varepsilon_{\kk,\tau}$ and $\Phi_{\kk,\tau}$ are respectively the fermion creation operation, moir\'e band energy, and wave function of the first conduction band state with spin label $s$, valley index $\tau$ and momentum $\kk$. Due to the time reversal symmetry, $\varepsilon_{\kk,\tau}=\varepsilon_{-\kk,-\tau}$ and $\Phi_{\kk,\tau}$=$\Phi^*_{-\kk,-\tau}$, where $\kk$ is measured relative to the moir\'e Brillouin zone center $\bar{\Gamma}$ point. In $H_1$, $A$ is the system area, $O_{\kk \kk'}^{(\tau)}(\qq)$ is the density matrix element, and $V(\qq)$ is the screened Coulomb potential $2\pi e^2 \tanh(q d)/(\epsilon q)$,
where $\epsilon$ is the effective dielectric constant, and $d$ is the vertical distance between TBG and top(bottom) metallic gates. We take $d$ to be 40 nm as in the experiment of Ref.~\cite{serlin2019intrinsic}, and $\epsilon$ as a free parameter since  screening  in TBG can be quite complicated. The dielectric screening from the encapsulating hBN should set a lower bound on $\epsilon$, leading to $\epsilon>5$ in our model. $\epsilon$ also effectively controls the ratio between interaction and bandwidth. In TBG the bandwidth near the magic angle is not exactly known experimentally, which is another good reason to take $\epsilon$ as a free parameter.

Hamiltonian $H$ has spin SU(2) and valley U(1) symmetry. We use the Hartree-Fock (HF) approximation and assume that the valley U(1) symmetry is preserved, but allow spin and valley polarization, which leads to the following mean-field Hamiltonian
\begin{equation}
\begin{aligned}
H_{\text{MF}} =& \sum_{\kk, \tau, s} E_{\kk,\tau, s} c^{\dagger}_{\kk,\tau, s} c_{\kk,\tau, s},\\
E_{\kk,\tau, s} =& \varepsilon_{\kk,\tau}+\frac{1}{A} \sum_{\kk', \tau', s'} V_{\kk \kk' \kk' \kk}^{(\tau \tau')} n_F(E_{\kk',\tau', s'})\\
 &-\frac{1}{A} \sum_{\kk'} V_{\kk \kk' \kk \kk'}^{(\tau \tau)} n_F(E_{\kk',\tau, s}),
\end{aligned}
\label{HFM}
\end{equation}
where the quasiparticle energy $E_{\kk,\tau, s}$ includes moir\'e band energy, and Hartree as well as Fock self energies, and $n_F$ is the Fermi-Dirac occupation number.

\begin{figure}[t!]
	\includegraphics[width=1\columnwidth]{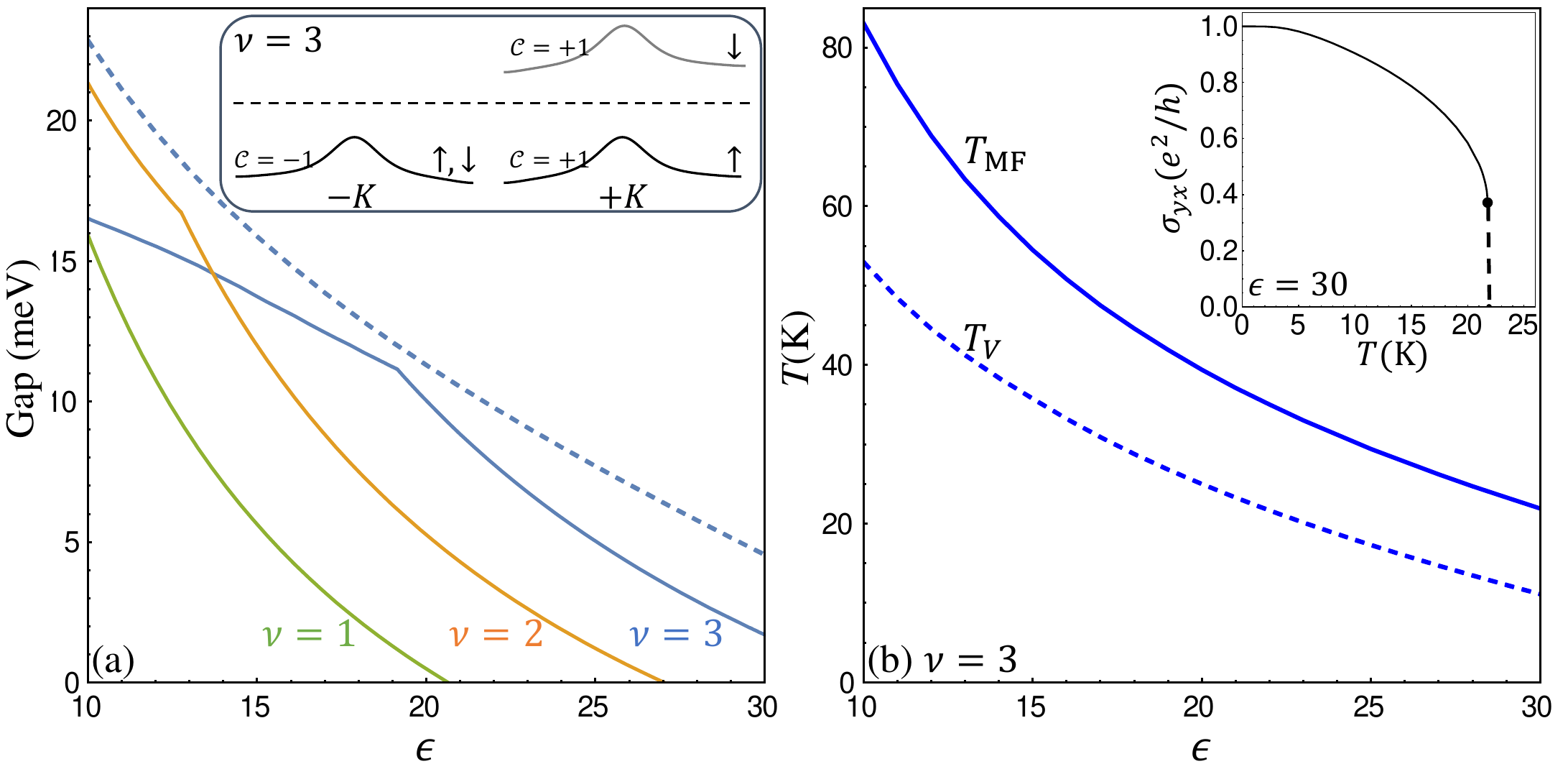}
	\caption{(a)Charged excitation gap as a function of dielectric constant $\epsilon$. The solid lines are the HF gap $\Delta_{\text{HF}}$ respectively for the three integer filling factors. The dashed line is the skyrmion-antiskyrmion pair energy $\Delta_{\text{pair}}$ at $\nu=3$. The inset schematically illustrate the QAHF at $\nu=3$. (b) Transition temperature at $\nu=3$ as a function of $\epsilon$. The solid line shows the mean-field transition temperature $T_{\text{MF}}$, and the dashed line shows the valley ordering temperature $T_V$ estimated using the valley wave spectrum. The inset presents the mean-field value of the anomalous Hall conductivity $\sigma_{yx}$, where the dashed line marks a jump in $\sigma_{yx}$ at $T_{\text{MF}}$. All calculations with interaction effects are done on a $36\times 36$ $k$-mesh. }
	\label{Fig:Tc}
\end{figure}

We focus on integer filling factors $\nu=$1, 2 and 3, and make a zero-temperature ($T=0$) ground state ansatz that $\nu$ out of the 4 first moir\'e conduction bands (including spin and valley degeneracies) are  filled, while the remaining $4-\nu$ bands are  empty.  At $\nu=1$ and 3, the ansatz leads to  maximally spin and valley  polarized states, which are QAHF and also exact eigenstates of the Hamiltonian $H$. At $\nu=2$, this ansatz generates two distinct types of states, namely, valley polarized state with QAHE and valley unpolarized state without QAHE, which are energetically degenerate  at this particular filling, but it is conceivable that a short-range atomic scale interaction (not included in our Hamiltonian $H$) may break this degeneracy. We calculate the $T=0$ HF energy gap $\Delta_{\text{HF}}$ between empty and occupied bands, as shown in Fig.~\ref{Fig:Tc}(a). A positive $\Delta_{\text{HF}}$ indicates the above ansatz is a good candidate for ground states at least in the HF approximation. As expected,  $\Delta_{\text{HF}}$ decreases with increasing dielectric constant $\epsilon$ because of the decreasing interaction strength. $\Delta_{\text{HF}}$ has a strong filling factor dependence, mainly because the Hartree self energy varies strongly with the electron density \cite{guinea2018electrostatic}. The gap $\Delta_{\text{HF}}$ at $\nu=3$ is smaller compared to those at $\nu=1$ and 2 for small $\epsilon$, but this order is reversed for large $\epsilon$.  By fitting to the  experimental $\nu=3$ gap ($\sim$ 2 meV) reported in Ref.~\cite{serlin2019intrinsic}, we estimate $\epsilon$ to be about 30 in our model. With this value of $\epsilon$, we find that  the $\nu=1$ and 2 states are not fully gapped in contrast to $\nu=3$, which is consistent with experimental findings in  Ref.~\cite{serlin2019intrinsic}. Therefore, our minimal model does capture the essential experimental phenomenology \cite{serlin2019intrinsic} provided $\epsilon$ is tuned to simulate screening of Coulomb interaction, most likely by all the other moir\'e bands neglected in our theory. 

We show the calculated mean-field ferromagnetic transition temperature $T_{\text{MF}}$ at $\nu=3$ in Fig.~\ref{Fig:Tc}(b).  $T_{\text{MF}}$($\epsilon=30$) is about 22 K, which is larger than the experimental Curie temperature ($\sim$9 K) \cite{serlin2019intrinsic}. We argue that this discrepancy is due to valley wave excitations, which limit the valley ordering temperature, as will be discussed in the following. The anomalous Hall conductivity $\sigma_{yx}$ at $\nu=3$ is quantized to $e^2/h$ within a 0.3\% accuracy up to $T=3 \text{K}$ as shown in Fig.~\ref{Fig:Tc}(b). We numerically find that $T_{\text{MF}}$ marks a first-order transition between phases with and without spin-valley polarization, which leads to a jump in $\sigma_{yx}$ at $T_{\text{MF}}$ [Fig.~\ref{Fig:Tc}(b)]. Remarkably, the experimental anomalous Hall resistance $R_{xy}$ in Ref.~\onlinecite{serlin2019intrinsic} also displays a sizable jump near the Curie temperature.

\begin{figure}[t!]
	\includegraphics[width=1\columnwidth]{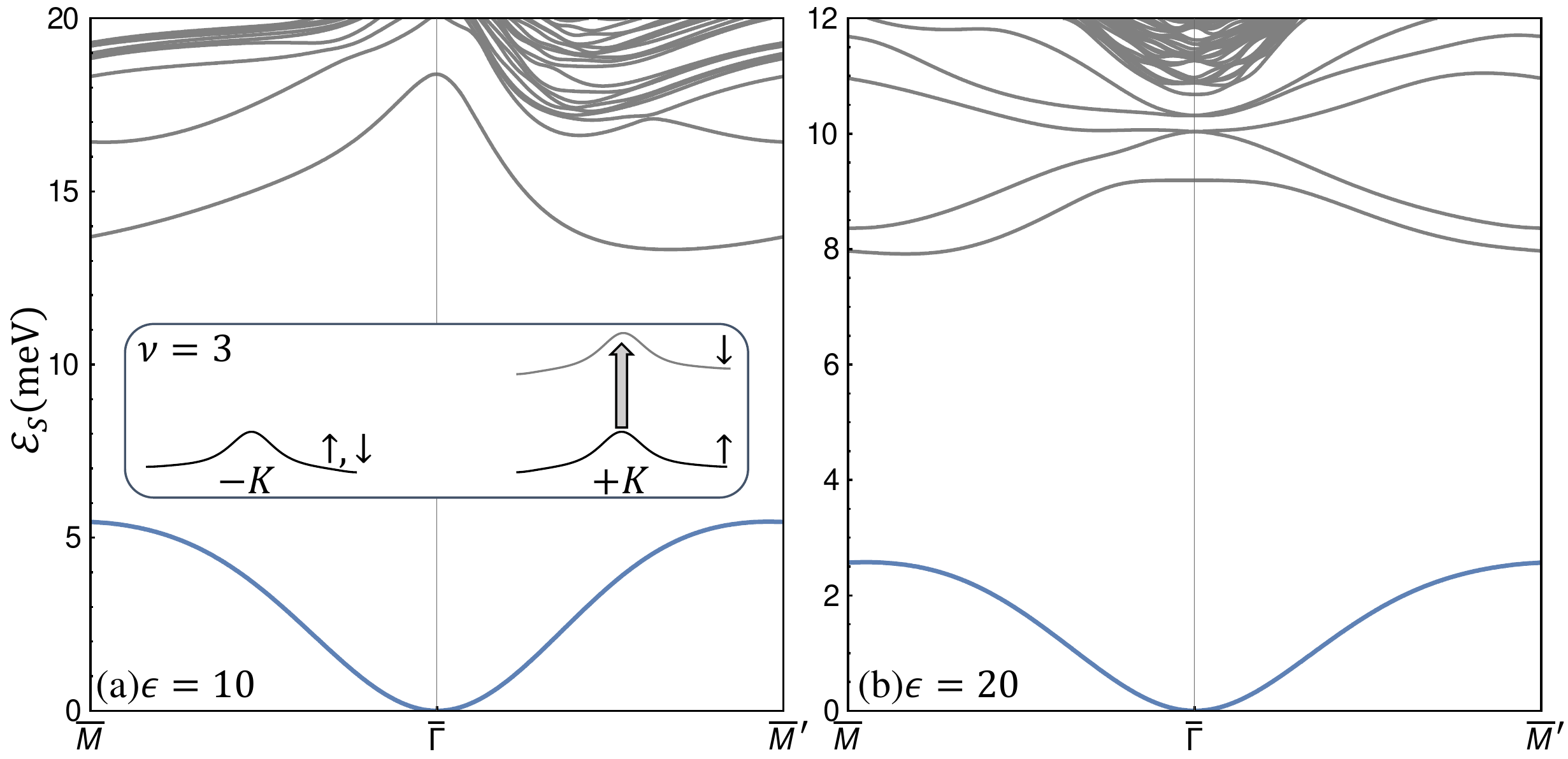}
	\caption{Excitation spectrum for $\nu=3$ spin magnon states [inset in (a) for illustration]. The blue lines in (a) and (b) represent the gapless spin wave mode. }
	\label{Fig:swave}
\end{figure}

{\it Spin wave.---} We examine the stability of the QAHF by studying the collective excitation spectrum. The spin magnon state at $\nu=3$ can be parametrized as follows
\begin{equation}
|\QQ\rangle_S = \sum_{\kk} z_{\kk,\QQ} c_{\kk+\QQ,+,\downarrow}^{\dagger} c_{\kk,+,\uparrow} |\nu=3\rangle
\end{equation}
where $|\nu=3\rangle$ is the QAHF state in which only the valley $+K$ and  spin $\downarrow$  band is empty, $z_{\kk,\QQ}$ are variational parameters, and $\QQ$ defined within the first moir\'e Brilouin zone is the momentum of the magnon. In the magnon state $|\QQ\rangle_S$, we make a single spin flip from the occupied  spin $\uparrow$  band to unoccupied spin $\downarrow$  band within the same $+K$ valley. Variation of the magnon energy with respect to $z_{\kk,\QQ}$ leads to the following eigenvalue problem
\begin{equation}
\begin{aligned}
&\mathcal{E}_S(\QQ)z_{\kk,\QQ}=\sum_{\kk'} \mathcal{H}_{\kk \kk'}^{(\QQ)}z_{\kk',\QQ},\\
&\mathcal{H}_{\kk \kk'}^{(\QQ)}= (E_{\kk+\QQ,+,\downarrow}-E_{\kk,+,\uparrow})\delta_{\kk,\kk'}-\frac{1}{A}V^{(++)}_{\kk' (\kk+\QQ) (\kk'+\QQ) \kk},
\end{aligned}
\label{SpinWave}
\end{equation}
where the first part in $\mathcal{H}_{\kk \kk'}^{(\QQ)}$ is the quasiparticle energy cost of the particle-hole transition, and the second part represents the electron-hole attraction.  Equation~(\ref{SpinWave}) is typically called the Bethe-Salpeter equation, representing repeated electron-hole interactions (``ladder diagrams''), in the context of excitons in semiconductors; here it gives rise to the spin wave spectrum. We note that $\mathcal{H}_{\kk \kk'}^{(\QQ)}$ is  not gauge invariant (except at $\QQ$=0) due to the phase ambiguity of the wave function. However, only closed loops in the momentum space appear in the characteristic polynomial of  $\mathcal{H}_{\kk \kk'}^{(\QQ)}$, making its eigenvalues gauge invariant; products of wave function overlap along the closed loops encode information of Berry curvature and quantum geometry \cite{Srivastava2015}.

\begin{figure}[t!]
	\includegraphics[width=1\columnwidth]{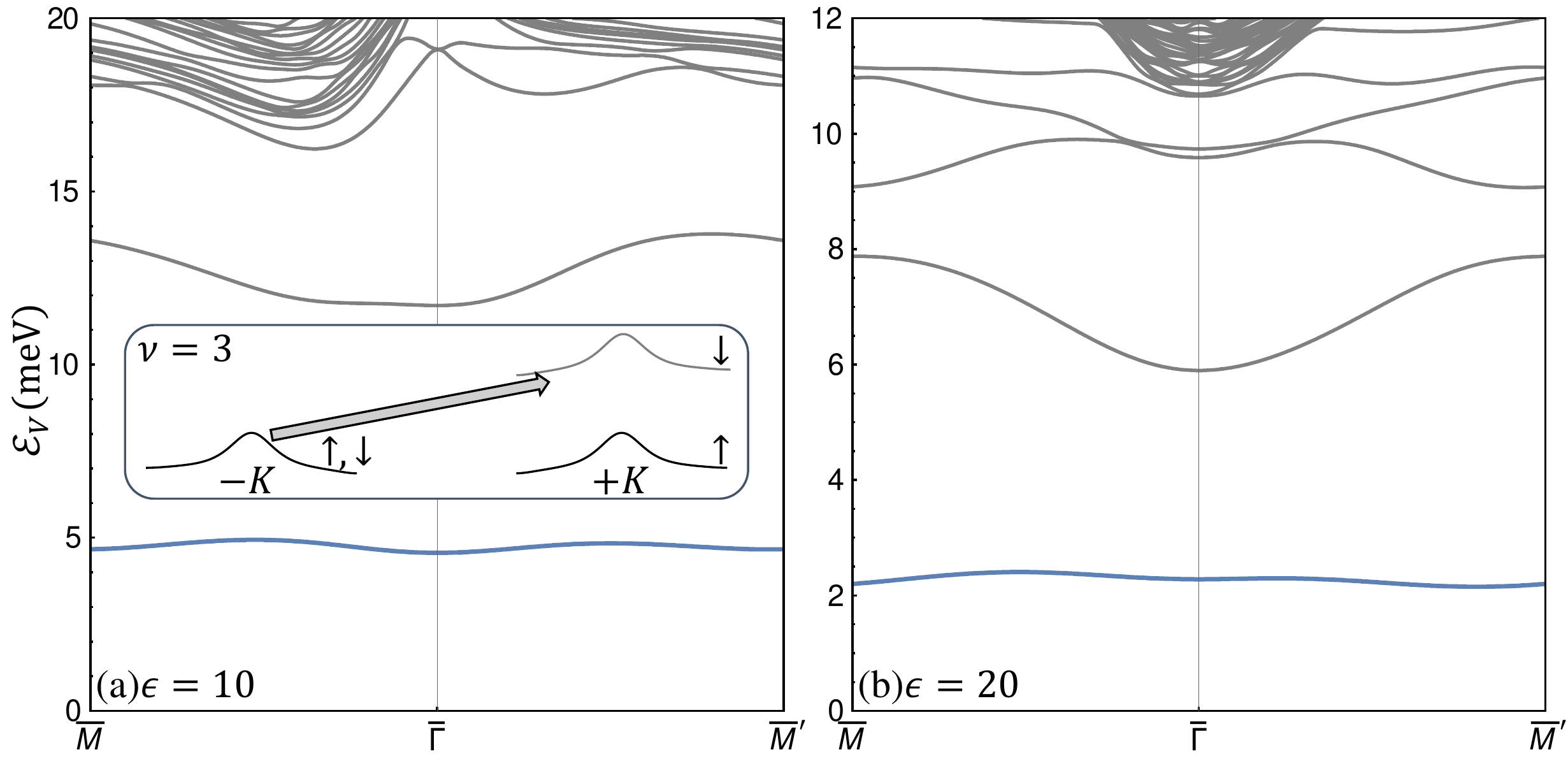}
	\caption{Excitation spectrum for $\nu=3$  valley magnon states [inset in (a) for illustration]. The blue lines in (a) and (b) represent the gapped valley wave mode.}
	\label{Fig:vwave}
\end{figure}

We numerically solve Eq.~(\ref{SpinWave}), and show the spin excitation spectrum in Fig.~\ref{Fig:swave}. The lowest energy mode (spin wave) is gapless at $\QQ=0$, which is expected from the Goldstone's theorem, as the continuous spin SU(2) symmetry is spontaneously broken in the QAHF. Because of the spin SU(2) symmetry, the spin lowering operator $\sum_{\kk}  c_{\kk,\tau,\downarrow}^{\dagger} c_{\kk,\tau,\uparrow}$ commutes with the Hamiltonian $H$. Therefore,  $z_{\kk,\QQ}=1$ for any $\kk$ is an exact zero-energy solution to Eq.~(\ref{SpinWave}) at $\QQ=0$.  The overall spin excitation spectrum is nonnegative in the parameter space that we have explored ($\epsilon$ up to 30), showing the stability of the QAHF at $\nu=3$ against spin wave excitations. 

The spin wave mode can be phenomenologically described using an $O$(3) nonlinear sigma model \cite{girvin1999quantum}
\begin{equation}
\mathcal{L}_S=-\int d^2 \rr \Big\{\frac{\hbar n_0}{2} \boldsymbol{\mathcal{A}}[\mm]\cdot \partial_t \mm+\frac{\rho_s}{2} (\boldsymbol{\nabla} \mm)^2 \Big\},
\label{LS}
\end{equation}
where the unit vector $\mm$ represents the local spin polarization, $\mathcal{A}[\mm]$ is the effective spin gauge field defined by $\boldsymbol{\nabla}_{\mm} \times \mathcal{A}[\mm] =\mm$, and $\rho_s$ the spin stiffness. We estimate $\rho_s$ by fitting the numerical spin wave spectrum around $\QQ=0$ shown in Fig.~\ref{Fig:swave} to the analytical spin wave dispersion $\mathcal{E}_{SW}=(2\rho_s/n_0) \QQ^2$ given by Eq.~(\ref{LS}). Besides spin waves, the Lagrangian $\mathcal{L}_S$ also supports skyrmion excitations, which are expected to be charged in the case of QAHF, similar to the QHF case \cite{Moon1995}. A pair of skyrmion and antiskyrmion has a total energy cost of $\Delta_{\text{pair}}=8\pi \rho_s$. We calculate $\Delta_{\text{pair}}$ using $\rho_s$ estimated above, and find that $\Delta_{\text{pair}}$ is comparable in magnitude to $\Delta_{\text{HF}}$,  but the former is larger at $\nu=3$, as shown in Fig.~\ref{Fig:Tc}(a). We find that the same ($\Delta_{\text{pair}}>\Delta_{\text{HF}}$) is also true at $\nu=1$ and 2 for spin maximally polarized states. By comparison, $\Delta_{\text{pair}}$ is half of $\Delta_{\text{HF}}$ for the $\nu=1$ quantum Hall ferromagnet in the lowest Landau level with Coulomb interaction \cite{Moon1995}.  An important difference here with the lowest Landau level  is that electron density in the moir\'e band is spatially nonuniform with modulation within the moir\'e unit cell, and both the Hartree and Fock self energies modify the moir\'e bandwidth. Nevertheless, we find that $\Delta_{\text{pair}}$ can be tuned to be smaller than $\Delta_{\text{HF}}$ in TBG by taking both  $\Delta_b$ and $\Delta_t$ to be finite (30 meV), which can be realized when both the top and bottom encapsulating hBN layers are in close alignment to TBG (see SM \cite{SM} for details).  Therefore, we conclude that the lowest charged excitation is determined by either $\Delta_{\text{HF}}$ or $\Delta_{\text{pair}}$, depending on system details.

{\it Valley wave.---} Besides spin magnon states, there are also valley magnon states with a single valley flip 
\begin{equation}
|\QQ\rangle_V = \sum_{\kk} z_{\kk,\QQ} c_{\kk+\QQ,+,\downarrow}^{\dagger} c_{\kk,-,s} |\nu=3\rangle,
\end{equation}
where $s$ can be either $\uparrow$ or $\downarrow$, since both spin components in $-K$ valley are fully occupied in $|\nu=3\rangle$. States $|\QQ\rangle_V$ with $s=\uparrow$ and $\downarrow$ are energetically degenerate for the Hamiltonian $H$, because it actually has an enlarged spin SU(2)$\times$SU(2) symmetry (independent spin rotation within each valley). The corresponding Bethe-Salpeter equation  is given by
\begin{equation}
\begin{aligned}
&\mathcal{E}_V(\QQ)z_{\kk,\QQ}=\sum_{\kk'} \mathcal{W}_{\kk \kk'}^{(\QQ)}z_{\kk',\QQ},\\
&\mathcal{W}_{\kk \kk'}^{(\QQ)}= (E_{\kk+\QQ,+,\downarrow}-E_{\kk,-,s})\delta_{\kk,\kk'}-\frac{1}{A}V^{(-+)}_{\kk' (\kk+\QQ) (\kk'+\QQ) \kk},
\end{aligned}
\end{equation}
which leads to the valley excitation spectrum in Fig.~\ref{Fig:vwave}. In contrast to the spin excitation spectrum, the lowest valley excitation mode (valley wave) is {\it gapped}, consistent with the fact that there is no continuous symmetry broken in the valley pseudospin space. The positive-energy valley wave  indicates  the robustness of $\nu=3$ QAHF against small variation in the valley space, which implies that the valley polarized state is energetically more favorable than the valley coherent state \cite{bultinck2019anomalous}. The valley wave can again be described by a nonlinear sigma model but with an Ising anisotropy 
\begin{equation}
\begin{aligned}
\mathcal{L}_V=-\int d^2 \rr \Big\{&\frac{\hbar n_0}{2} \boldsymbol{\mathcal{A}}[\PP]\cdot \partial_t \PP-u\pi_z^2+\frac{\rho_z}{2} (\boldsymbol{\nabla} \pi_z)^2 \\
&+\frac{\rho_\perp}{2} [(\boldsymbol{\nabla} \pi_x)^2+(\boldsymbol{\nabla} \pi_y)^2] \Big\},
\end{aligned}
\label{LV}
\end{equation}
where the unit vector $\PP$ represents the local valley polarization ($\pi_z$ for valley Ising order and $\pi_{x,y}$ for valley coherent order), $u>0$ captures the Ising anisotropy, $\rho_{z, \perp}$ are anisotropic valley stiffness, and other terms are similar to those in Eq.~(\ref{LS}). The analytical valley wave dispersion is
$\mathcal{E}_{VW}=\Delta_V+(2\rho_\perp/n_0)\QQ^2$, where $\Delta_V=4 u/n_0$. Therefore, we can estimate $u$ and $\rho_\perp$ using the numerical valley excitation spectrum in Fig.~\ref{Fig:vwave}. 

Because of the Ising anisotropy, there can be valley domain excitations. We make a domain wall ansatz $(\pi_x,\pi_y,\pi_z)=[\text{sech}(x/\lambda),0,\tanh(x/\lambda)]$, and its energy cost is minimized by taking the domain wall width $\lambda$ to be $\sqrt{(\rho_\perp+2\rho_z)/(6u)}$. The domain wall energy per length is then $J=4 u \lambda$. We note that this domain wall separates regions with opposite Chern numbers, and binds one-dimensional chiral electronic states. The valley Ising ordering temperature limited by the proliferation of domain walls can be estimated to be \cite{Chaiken,XiaoLi2014}
\begin{equation}
k_B T_{DW} =\frac{2}{\ln(1+\sqrt{2})} J \lambda \approx 2.62 \Big(\frac{\lambda}{a_M}\Big)^2 \Delta_{V},
\end{equation}
where $a_M$ is the moir\'e period. $\Delta_{V}$ can be directly extracted from the valley wave spectrum, but $\lambda$ can not because $\mathcal{E}_{VW}$ has no dependence on $\rho_z$. Since $a_M$ is the lattice scale in our problem, we argue that the domain wall width $\lambda$ is larger than $a_M$, and therefore, we estimate that $k_B T_{DW}>2.62 \Delta_{V}$. On the other hand, valley waves are already thermally excited when $k_B T$ exceeds $\Delta_{V}$. Therefore, we conclude that the valley ordering temperature $T_V$ is mostly limited by valley waves instead of domain walls, and estimate $k_B T_V$ from the valley wave minimum  energy. The resulting $T_V$ is shown in Fig.~\ref{Fig:Tc}(b), which is below the mean-field transition temperature $T_{\text{MF}}$. For a zero-temperature charged excitation gap of 2 meV, we find a corresponding $T_V$ of about 11 K, which compares well with the experimental Curie temperature \cite{serlin2019intrinsic}. Although this good quantitative agreement with experiment might be a coincidence, our work establishes the emergent TBG QAHF to be likely a valley Ising ordered state.  Regarding the finite jump in the experimentally measured $R_{xy}$ near the transition temperature \cite{serlin2019intrinsic}, we provide a possible theoretical scenario that the interplay between the continuous spin order parameter $\boldsymbol{m}$ and the Ising degree-of-freedom $\pi_z$ through higher-order coupling terms (not included in $\mathcal{L}_S$ and $\mathcal{L}_V$) could change the finite-temperature phase transition from second order to first order \cite{Kato2010}.

{\it Discussion.---} In summary, we present a microscopic theory for spin and valley waves of QAHF in TBG, and demonstrate that the excitation spectra provide important information about the stability of mean-field state, the transport energy gap, and the valley ordering temperature. We find that TBG QAHF is robust, provided that other effects such as disorder can be neglected. Besides ferromagnetism, flat moir\'e bands can host a rich set of broken symmetry states. Our theory can be generalized to study collective excitations of other broken symmetry states in moir\'e materials.

%\section{acknowledgment}
F. W. thanks A. Young,  M. Zaletel, N. Bultinck, S. Chatterjee and I. Martin for discussions. This work was initiated at the Aspen Center for Physics, which is supported by National Science Foundation grant PHY-1607611. We acknowledge support by the Laboratory for Physical Sciences.

{\it Note added.} While this paper was being written, three related arXiv preprints \cite{repellin2019ferromagnetism,alavirad2019ferromagnetism,chatterjee2019symmetry} appeared. In this paper we addressed valley ordering temperature limited by valley wave excitations, which has not been studied previously in TBG to our knowledge.

\bibliographystyle{apsrev4-1}
\bibliography{refs}

\begin{thebibliography}{76}%
\makeatletter
\providecommand \@ifxundefined [1]{%
 \@ifx{#1\undefined}
}%
\providecommand \@ifnum [1]{%
 \ifnum #1\expandafter \@firstoftwo
 \else \expandafter \@secondoftwo
 \fi
}%
\providecommand \@ifx [1]{%
 \ifx #1\expandafter \@firstoftwo
 \else \expandafter \@secondoftwo
 \fi
}%
\providecommand \natexlab [1]{#1}%
\providecommand \enquote  [1]{``#1''}%
\providecommand \bibnamefont  [1]{#1}%
\providecommand \bibfnamefont [1]{#1}%
\providecommand \citenamefont [1]{#1}%
\providecommand \href@noop [0]{\@secondoftwo}%
\providecommand \href [0]{\begingroup \@sanitize@url \@href}%
\providecommand \@href[1]{\@@startlink{#1}\@@href}%
\providecommand \@@href[1]{\endgroup#1\@@endlink}%
\providecommand \@sanitize@url [0]{\catcode `\\12\catcode `\$12\catcode
  `\&12\catcode `\#12\catcode `\^12\catcode `\_12\catcode `\%12\relax}%
\providecommand \@@startlink[1]{}%
\providecommand \@@endlink[0]{}%
\providecommand \url  [0]{\begingroup\@sanitize@url \@url }%
\providecommand \@url [1]{\endgroup\@href {#1}{\urlprefix }}%
\providecommand \urlprefix  [0]{URL }%
\providecommand \Eprint [0]{\href }%
\providecommand \doibase [0]{http://dx.doi.org/}%
\providecommand \selectlanguage [0]{\@gobble}%
\providecommand \bibinfo  [0]{\@secondoftwo}%
\providecommand \bibfield  [0]{\@secondoftwo}%
\providecommand \translation [1]{[#1]}%
\providecommand \BibitemOpen [0]{}%
\providecommand \bibitemStop [0]{}%
\providecommand \bibitemNoStop [0]{.\EOS\space}%
\providecommand \EOS [0]{\spacefactor3000\relax}%
\providecommand \BibitemShut  [1]{\csname bibitem#1\endcsname}%
\let\auto@bib@innerbib\@empty
%</preamble>
\bibitem [{\citenamefont {Cao}\ \emph {et~al.}(2018{\natexlab{a}})\citenamefont
  {Cao}, \citenamefont {Fatemi}, \citenamefont {Fang}, \citenamefont
  {Watanabe}, \citenamefont {Taniguchi}, \citenamefont {Kaxiras},\ and\
  \citenamefont {Jarillo-Herrero}}]{Cao2018Super}%
  \BibitemOpen
  \bibfield  {author} {\bibinfo {author} {\bibfnamefont {Y.}~\bibnamefont
  {Cao}}, \bibinfo {author} {\bibfnamefont {V.}~\bibnamefont {Fatemi}},
  \bibinfo {author} {\bibfnamefont {S.}~\bibnamefont {Fang}}, \bibinfo {author}
  {\bibfnamefont {K.}~\bibnamefont {Watanabe}}, \bibinfo {author}
  {\bibfnamefont {T.}~\bibnamefont {Taniguchi}}, \bibinfo {author}
  {\bibfnamefont {E.}~\bibnamefont {Kaxiras}}, \ and\ \bibinfo {author}
  {\bibfnamefont {P.}~\bibnamefont {Jarillo-Herrero}},\ }\href
  {http://dx.doi.org/10.1038/nature26160} {\bibfield  {journal} {\bibinfo
  {journal} {Nature}\ }\textbf {\bibinfo {volume} {556}},\ \bibinfo {pages}
  {43} (\bibinfo {year} {2018}{\natexlab{a}})}\BibitemShut {NoStop}%
\bibitem [{\citenamefont {Cao}\ \emph {et~al.}(2018{\natexlab{b}})\citenamefont
  {Cao}, \citenamefont {Fatemi}, \citenamefont {Demir}, \citenamefont {Fang},
  \citenamefont {Tomarken}, \citenamefont {Luo}, \citenamefont
  {Sanchez-Yamagishi}, \citenamefont {Watanabe}, \citenamefont {Taniguchi},
  \citenamefont {Kaxiras}, \citenamefont {Ashoori},\ and\ \citenamefont
  {Jarillo-Herrero}}]{Cao2018Magnetic}%
  \BibitemOpen
  \bibfield  {author} {\bibinfo {author} {\bibfnamefont {Y.}~\bibnamefont
  {Cao}}, \bibinfo {author} {\bibfnamefont {V.}~\bibnamefont {Fatemi}},
  \bibinfo {author} {\bibfnamefont {A.}~\bibnamefont {Demir}}, \bibinfo
  {author} {\bibfnamefont {S.}~\bibnamefont {Fang}}, \bibinfo {author}
  {\bibfnamefont {S.~L.}\ \bibnamefont {Tomarken}}, \bibinfo {author}
  {\bibfnamefont {J.~Y.}\ \bibnamefont {Luo}}, \bibinfo {author} {\bibfnamefont
  {J.~D.}\ \bibnamefont {Sanchez-Yamagishi}}, \bibinfo {author} {\bibfnamefont
  {K.}~\bibnamefont {Watanabe}}, \bibinfo {author} {\bibfnamefont
  {T.}~\bibnamefont {Taniguchi}}, \bibinfo {author} {\bibfnamefont
  {E.}~\bibnamefont {Kaxiras}}, \bibinfo {author} {\bibfnamefont {R.~C.}\
  \bibnamefont {Ashoori}}, \ and\ \bibinfo {author} {\bibfnamefont
  {P.}~\bibnamefont {Jarillo-Herrero}},\ }\href
  {http://dx.doi.org/10.1038/nature26154} {\bibfield  {journal} {\bibinfo
  {journal} {Nature}\ }\textbf {\bibinfo {volume} {556}},\ \bibinfo {pages}
  {80} (\bibinfo {year} {2018}{\natexlab{b}})}\BibitemShut {NoStop}%
\bibitem [{\citenamefont {Kerelsky}\ \emph {et~al.}(2019)\citenamefont
  {Kerelsky}, \citenamefont {McGilly}, \citenamefont {Kennes}, \citenamefont
  {Xian}, \citenamefont {Yankowitz}, \citenamefont {Chen}, \citenamefont
  {Watanabe}, \citenamefont {Taniguchi}, \citenamefont {Hone}, \citenamefont
  {Dean}, \citenamefont {Rubio},\ and\ \citenamefont
  {Pasupathy}}]{Kerelsky2019}%
  \BibitemOpen
  \bibfield  {author} {\bibinfo {author} {\bibfnamefont {A.}~\bibnamefont
  {Kerelsky}}, \bibinfo {author} {\bibfnamefont {L.~J.}\ \bibnamefont
  {McGilly}}, \bibinfo {author} {\bibfnamefont {D.~M.}\ \bibnamefont {Kennes}},
  \bibinfo {author} {\bibfnamefont {L.}~\bibnamefont {Xian}}, \bibinfo {author}
  {\bibfnamefont {M.}~\bibnamefont {Yankowitz}}, \bibinfo {author}
  {\bibfnamefont {S.}~\bibnamefont {Chen}}, \bibinfo {author} {\bibfnamefont
  {K.}~\bibnamefont {Watanabe}}, \bibinfo {author} {\bibfnamefont
  {T.}~\bibnamefont {Taniguchi}}, \bibinfo {author} {\bibfnamefont
  {J.}~\bibnamefont {Hone}}, \bibinfo {author} {\bibfnamefont {C.}~\bibnamefont
  {Dean}}, \bibinfo {author} {\bibfnamefont {A.}~\bibnamefont {Rubio}}, \ and\
  \bibinfo {author} {\bibfnamefont {A.~N.}\ \bibnamefont {Pasupathy}},\ }\href
  {https://doi.org/10.1038/s41586-019-1431-9} {\bibfield  {journal} {\bibinfo
  {journal} {Nature}\ }\textbf {\bibinfo {volume} {572}},\ \bibinfo {pages}
  {95} (\bibinfo {year} {2019})}\BibitemShut {NoStop}%
\bibitem [{\citenamefont {Choi}\ \emph {et~al.}()\citenamefont {Choi},
  \citenamefont {Kemmer}, \citenamefont {Peng}, \citenamefont {Thomson},
  \citenamefont {Arora}, \citenamefont {Polski}, \citenamefont {Zhang},
  \citenamefont {Ren}, \citenamefont {Alicea}, \citenamefont {Refael} \emph
  {et~al.}}]{choi2019imaging}%
  \BibitemOpen
  \bibfield  {author} {\bibinfo {author} {\bibfnamefont {Y.}~\bibnamefont
  {Choi}}, \bibinfo {author} {\bibfnamefont {J.}~\bibnamefont {Kemmer}},
  \bibinfo {author} {\bibfnamefont {Y.}~\bibnamefont {Peng}}, \bibinfo {author}
  {\bibfnamefont {A.}~\bibnamefont {Thomson}}, \bibinfo {author} {\bibfnamefont
  {H.}~\bibnamefont {Arora}}, \bibinfo {author} {\bibfnamefont
  {R.}~\bibnamefont {Polski}}, \bibinfo {author} {\bibfnamefont
  {Y.}~\bibnamefont {Zhang}}, \bibinfo {author} {\bibfnamefont
  {H.}~\bibnamefont {Ren}}, \bibinfo {author} {\bibfnamefont {J.}~\bibnamefont
  {Alicea}}, \bibinfo {author} {\bibfnamefont {G.}~\bibnamefont {Refael}},
  \emph {et~al.},\ }\href {https://arxiv.org/abs/1901.02997} {\bibinfo
  {journal} {arXiv:1901.02997}\ }\BibitemShut {NoStop}%
\bibitem [{\citenamefont {Cao}\ \emph {et~al.}({\natexlab{a}})\citenamefont
  {Cao}, \citenamefont {Chowdhury}, \citenamefont {Rodan-Legrain},
  \citenamefont {Rubies-Bigord{\`a}}, \citenamefont {Watanabe}, \citenamefont
  {Taniguchi}, \citenamefont {Senthil},\ and\ \citenamefont
  {Jarillo-Herrero}}]{MIT2018_rho}%
  \BibitemOpen
\bibfield  {journal} {  }\bibfield  {author} {\bibinfo {author} {\bibfnamefont
  {Y.}~\bibnamefont {Cao}}, \bibinfo {author} {\bibfnamefont {D.}~\bibnamefont
  {Chowdhury}}, \bibinfo {author} {\bibfnamefont {D.}~\bibnamefont
  {Rodan-Legrain}}, \bibinfo {author} {\bibfnamefont {O.}~\bibnamefont
  {Rubies-Bigord{\`a}}}, \bibinfo {author} {\bibfnamefont {K.}~\bibnamefont
  {Watanabe}}, \bibinfo {author} {\bibfnamefont {T.}~\bibnamefont {Taniguchi}},
  \bibinfo {author} {\bibfnamefont {T.}~\bibnamefont {Senthil}}, \ and\
  \bibinfo {author} {\bibfnamefont {P.}~\bibnamefont {Jarillo-Herrero}},\
  }\href {https://arxiv.org/abs/1901.03710} {\bibfield  {journal} {\bibinfo
  {journal} {arXiv:1901.03710}\ } ({\natexlab{a}})}\BibitemShut {NoStop}%
\bibitem [{\citenamefont {Polshyn}\ \emph {et~al.}()\citenamefont {Polshyn},
  \citenamefont {Yankowitz}, \citenamefont {Chen}, \citenamefont {Zhang},
  \citenamefont {Watanabe}, \citenamefont {Taniguchi}, \citenamefont {Dean},\
  and\ \citenamefont {Young}}]{Columbia2018_rho}%
  \BibitemOpen
  \bibfield  {author} {\bibinfo {author} {\bibfnamefont {H.}~\bibnamefont
  {Polshyn}}, \bibinfo {author} {\bibfnamefont {M.}~\bibnamefont {Yankowitz}},
  \bibinfo {author} {\bibfnamefont {S.}~\bibnamefont {Chen}}, \bibinfo {author}
  {\bibfnamefont {Y.}~\bibnamefont {Zhang}}, \bibinfo {author} {\bibfnamefont
  {K.}~\bibnamefont {Watanabe}}, \bibinfo {author} {\bibfnamefont
  {T.}~\bibnamefont {Taniguchi}}, \bibinfo {author} {\bibfnamefont {C.~R.}\
  \bibnamefont {Dean}}, \ and\ \bibinfo {author} {\bibfnamefont {A.~F.}\
  \bibnamefont {Young}},\ }\href {https://arxiv.org/abs/1902.00763} {\bibinfo
  {journal} {arXiv:1902.00763}\ }\BibitemShut {NoStop}%
\bibitem [{\citenamefont {Sharpe}\ \emph {et~al.}(2019)\citenamefont {Sharpe},
  \citenamefont {Fox}, \citenamefont {Barnard}, \citenamefont {Finney},
  \citenamefont {Watanabe}, \citenamefont {Taniguchi}, \citenamefont
  {Kastner},\ and\ \citenamefont {Goldhaber-Gordon}}]{sharpe2019emergent}%
  \BibitemOpen
\bibfield  {journal} {  }\bibfield  {author} {\bibinfo {author} {\bibfnamefont
  {A.~L.}\ \bibnamefont {Sharpe}}, \bibinfo {author} {\bibfnamefont {E.~J.}\
  \bibnamefont {Fox}}, \bibinfo {author} {\bibfnamefont {A.~W.}\ \bibnamefont
  {Barnard}}, \bibinfo {author} {\bibfnamefont {J.}~\bibnamefont {Finney}},
  \bibinfo {author} {\bibfnamefont {K.}~\bibnamefont {Watanabe}}, \bibinfo
  {author} {\bibfnamefont {T.}~\bibnamefont {Taniguchi}}, \bibinfo {author}
  {\bibfnamefont {M.~A.}\ \bibnamefont {Kastner}}, \ and\ \bibinfo {author}
  {\bibfnamefont {D.}~\bibnamefont {Goldhaber-Gordon}},\ }\href {\doibase
  10.1126/science.aaw3780} {\bibfield  {journal} {\bibinfo  {journal}
  {Science}\ }\textbf {\bibinfo {volume} {365}},\ \bibinfo {pages} {605}
  (\bibinfo {year} {2019})}\BibitemShut {NoStop}%
\bibitem [{\citenamefont {Serlin}\ \emph {et~al.}()\citenamefont {Serlin},
  \citenamefont {Tschirhart}, \citenamefont {Polshyn}, \citenamefont {Zhang},
  \citenamefont {Zhu}, \citenamefont {Watanabe}, \citenamefont {Taniguchi},
  \citenamefont {Balents},\ and\ \citenamefont {Young}}]{serlin2019intrinsic}%
  \BibitemOpen
  \bibfield  {author} {\bibinfo {author} {\bibfnamefont {M.}~\bibnamefont
  {Serlin}}, \bibinfo {author} {\bibfnamefont {C.}~\bibnamefont {Tschirhart}},
  \bibinfo {author} {\bibfnamefont {H.}~\bibnamefont {Polshyn}}, \bibinfo
  {author} {\bibfnamefont {Y.}~\bibnamefont {Zhang}}, \bibinfo {author}
  {\bibfnamefont {J.}~\bibnamefont {Zhu}}, \bibinfo {author} {\bibfnamefont
  {K.}~\bibnamefont {Watanabe}}, \bibinfo {author} {\bibfnamefont
  {T.}~\bibnamefont {Taniguchi}}, \bibinfo {author} {\bibfnamefont
  {L.}~\bibnamefont {Balents}}, \ and\ \bibinfo {author} {\bibfnamefont
  {A.}~\bibnamefont {Young}},\ }\href {https://arxiv.org/abs/1907.00261}
  {\bibinfo  {journal} {arXiv:1907.00261}\ }\BibitemShut {NoStop}%
\bibitem [{\citenamefont {Chen}\ \emph {et~al.}()\citenamefont {Chen},
  \citenamefont {Sharpe}, \citenamefont {Fox}, \citenamefont {Zhang},
  \citenamefont {Wang}, \citenamefont {Jiang}, \citenamefont {Lyu},
  \citenamefont {Li}, \citenamefont {Watanabe}, \citenamefont {Taniguchi} \emph
  {et~al.}}]{chen2019tunable}%
  \BibitemOpen
\bibfield  {journal} {  }\bibfield  {author} {\bibinfo {author} {\bibfnamefont
  {G.}~\bibnamefont {Chen}}, \bibinfo {author} {\bibfnamefont {A.~L.}\
  \bibnamefont {Sharpe}}, \bibinfo {author} {\bibfnamefont {E.~J.}\
  \bibnamefont {Fox}}, \bibinfo {author} {\bibfnamefont {Y.-H.}\ \bibnamefont
  {Zhang}}, \bibinfo {author} {\bibfnamefont {S.}~\bibnamefont {Wang}},
  \bibinfo {author} {\bibfnamefont {L.}~\bibnamefont {Jiang}}, \bibinfo
  {author} {\bibfnamefont {B.}~\bibnamefont {Lyu}}, \bibinfo {author}
  {\bibfnamefont {H.}~\bibnamefont {Li}}, \bibinfo {author} {\bibfnamefont
  {K.}~\bibnamefont {Watanabe}}, \bibinfo {author} {\bibfnamefont
  {T.}~\bibnamefont {Taniguchi}},  \emph {et~al.},\ }\href
  {https://arxiv.org/abs/1905.06535} {\bibinfo  {journal} {arXiv:1905.06535}\
  }\BibitemShut {NoStop}%
\bibitem [{\citenamefont {Yankowitz}\ \emph {et~al.}(2019)\citenamefont
  {Yankowitz}, \citenamefont {Chen}, \citenamefont {Polshyn}, \citenamefont
  {Zhang}, \citenamefont {Watanabe}, \citenamefont {Taniguchi}, \citenamefont
  {Graf}, \citenamefont {Young},\ and\ \citenamefont {Dean}}]{Dean2018tuning}%
  \BibitemOpen
\bibfield  {journal} {  }\bibfield  {author} {\bibinfo {author} {\bibfnamefont
  {M.}~\bibnamefont {Yankowitz}}, \bibinfo {author} {\bibfnamefont
  {S.}~\bibnamefont {Chen}}, \bibinfo {author} {\bibfnamefont {H.}~\bibnamefont
  {Polshyn}}, \bibinfo {author} {\bibfnamefont {Y.}~\bibnamefont {Zhang}},
  \bibinfo {author} {\bibfnamefont {K.}~\bibnamefont {Watanabe}}, \bibinfo
  {author} {\bibfnamefont {T.}~\bibnamefont {Taniguchi}}, \bibinfo {author}
  {\bibfnamefont {D.}~\bibnamefont {Graf}}, \bibinfo {author} {\bibfnamefont
  {A.~F.}\ \bibnamefont {Young}}, \ and\ \bibinfo {author} {\bibfnamefont
  {C.~R.}\ \bibnamefont {Dean}},\ }\href
  {https://science.sciencemag.org/content/363/6431/1059} {\bibfield  {journal}
  {\bibinfo  {journal} {Science}\ }\textbf {\bibinfo {volume} {363}},\ \bibinfo
  {pages} {1059} (\bibinfo {year} {2019})}\BibitemShut {NoStop}%
\bibitem [{\citenamefont {Codecido}\ \emph {et~al.}()\citenamefont {Codecido},
  \citenamefont {Wang}, \citenamefont {Koester}, \citenamefont {Che},
  \citenamefont {Tian}, \citenamefont {Lv}, \citenamefont {Tran}, \citenamefont
  {Watanabe}, \citenamefont {Taniguchi}, \citenamefont {Zhang} \emph
  {et~al.}}]{codecido2019correlated}%
  \BibitemOpen
  \bibfield  {author} {\bibinfo {author} {\bibfnamefont {E.}~\bibnamefont
  {Codecido}}, \bibinfo {author} {\bibfnamefont {Q.}~\bibnamefont {Wang}},
  \bibinfo {author} {\bibfnamefont {R.}~\bibnamefont {Koester}}, \bibinfo
  {author} {\bibfnamefont {S.}~\bibnamefont {Che}}, \bibinfo {author}
  {\bibfnamefont {H.}~\bibnamefont {Tian}}, \bibinfo {author} {\bibfnamefont
  {R.}~\bibnamefont {Lv}}, \bibinfo {author} {\bibfnamefont {S.}~\bibnamefont
  {Tran}}, \bibinfo {author} {\bibfnamefont {K.}~\bibnamefont {Watanabe}},
  \bibinfo {author} {\bibfnamefont {T.}~\bibnamefont {Taniguchi}}, \bibinfo
  {author} {\bibfnamefont {F.}~\bibnamefont {Zhang}},  \emph {et~al.},\ }\href
  {https://arxiv.org/abs/1902.05151} {\bibinfo  {journal} {arXiv:1902.05151}\
  }\BibitemShut {NoStop}%
\bibitem [{\citenamefont {Lu}\ \emph {et~al.}()\citenamefont {Lu},
  \citenamefont {Stepanov}, \citenamefont {Yang}, \citenamefont {Xie},
  \citenamefont {Aamir}, \citenamefont {Das}, \citenamefont {Urgell},
  \citenamefont {Watanabe}, \citenamefont {Taniguchi}, \citenamefont {Zhang}
  \emph {et~al.}}]{lu2019superconductors}%
  \BibitemOpen
\bibfield  {journal} {  }\bibfield  {author} {\bibinfo {author} {\bibfnamefont
  {X.}~\bibnamefont {Lu}}, \bibinfo {author} {\bibfnamefont {P.}~\bibnamefont
  {Stepanov}}, \bibinfo {author} {\bibfnamefont {W.}~\bibnamefont {Yang}},
  \bibinfo {author} {\bibfnamefont {M.}~\bibnamefont {Xie}}, \bibinfo {author}
  {\bibfnamefont {M.~A.}\ \bibnamefont {Aamir}}, \bibinfo {author}
  {\bibfnamefont {I.}~\bibnamefont {Das}}, \bibinfo {author} {\bibfnamefont
  {C.}~\bibnamefont {Urgell}}, \bibinfo {author} {\bibfnamefont
  {K.}~\bibnamefont {Watanabe}}, \bibinfo {author} {\bibfnamefont
  {T.}~\bibnamefont {Taniguchi}}, \bibinfo {author} {\bibfnamefont
  {G.}~\bibnamefont {Zhang}},  \emph {et~al.},\ }\href
  {https://arxiv.org/abs/1903.06513} {\bibinfo  {journal} {arXiv:1903.06513}\
  }\BibitemShut {NoStop}%
\bibitem [{\citenamefont {Tomarken}\ \emph {et~al.}(2019)\citenamefont
  {Tomarken}, \citenamefont {Cao}, \citenamefont {Demir}, \citenamefont
  {Watanabe}, \citenamefont {Taniguchi}, \citenamefont {Jarillo-Herrero},\ and\
  \citenamefont {Ashoori}}]{tomarken2019electronic}%
  \BibitemOpen
\bibfield  {journal} {  }\bibfield  {author} {\bibinfo {author} {\bibfnamefont
  {S.~L.}\ \bibnamefont {Tomarken}}, \bibinfo {author} {\bibfnamefont
  {Y.}~\bibnamefont {Cao}}, \bibinfo {author} {\bibfnamefont {A.}~\bibnamefont
  {Demir}}, \bibinfo {author} {\bibfnamefont {K.}~\bibnamefont {Watanabe}},
  \bibinfo {author} {\bibfnamefont {T.}~\bibnamefont {Taniguchi}}, \bibinfo
  {author} {\bibfnamefont {P.}~\bibnamefont {Jarillo-Herrero}}, \ and\ \bibinfo
  {author} {\bibfnamefont {R.~C.}\ \bibnamefont {Ashoori}},\ }\href {\doibase
  10.1103/PhysRevLett.123.046601} {\bibfield  {journal} {\bibinfo  {journal}
  {Phys. Rev. Lett.}\ }\textbf {\bibinfo {volume} {123}},\ \bibinfo {pages}
  {046601} (\bibinfo {year} {2019})}\BibitemShut {NoStop}%
\bibitem [{\citenamefont {Xie}\ \emph {et~al.}(2019)\citenamefont {Xie},
  \citenamefont {Lian}, \citenamefont {J{\"a}ck}, \citenamefont {Liu},
  \citenamefont {Chiu}, \citenamefont {Watanabe}, \citenamefont {Taniguchi},
  \citenamefont {Bernevig},\ and\ \citenamefont {Yazdani}}]{Xie2019}%
  \BibitemOpen
  \bibfield  {author} {\bibinfo {author} {\bibfnamefont {Y.}~\bibnamefont
  {Xie}}, \bibinfo {author} {\bibfnamefont {B.}~\bibnamefont {Lian}}, \bibinfo
  {author} {\bibfnamefont {B.}~\bibnamefont {J{\"a}ck}}, \bibinfo {author}
  {\bibfnamefont {X.}~\bibnamefont {Liu}}, \bibinfo {author} {\bibfnamefont
  {C.-L.}\ \bibnamefont {Chiu}}, \bibinfo {author} {\bibfnamefont
  {K.}~\bibnamefont {Watanabe}}, \bibinfo {author} {\bibfnamefont
  {T.}~\bibnamefont {Taniguchi}}, \bibinfo {author} {\bibfnamefont {B.~A.}\
  \bibnamefont {Bernevig}}, \ and\ \bibinfo {author} {\bibfnamefont
  {A.}~\bibnamefont {Yazdani}},\ }\href
  {https://doi.org/10.1038/s41586-019-1422-x} {\bibfield  {journal} {\bibinfo
  {journal} {Nature}\ }\textbf {\bibinfo {volume} {572}},\ \bibinfo {pages}
  {101} (\bibinfo {year} {2019})}\BibitemShut {NoStop}%
\bibitem [{\citenamefont {Jiang}\ \emph {et~al.}(2019)\citenamefont {Jiang},
  \citenamefont {Lai}, \citenamefont {Watanabe}, \citenamefont {Taniguchi},
  \citenamefont {Haule}, \citenamefont {Mao},\ and\ \citenamefont
  {Andrei}}]{Jiang2019}%
  \BibitemOpen
  \bibfield  {author} {\bibinfo {author} {\bibfnamefont {Y.}~\bibnamefont
  {Jiang}}, \bibinfo {author} {\bibfnamefont {X.}~\bibnamefont {Lai}}, \bibinfo
  {author} {\bibfnamefont {K.}~\bibnamefont {Watanabe}}, \bibinfo {author}
  {\bibfnamefont {T.}~\bibnamefont {Taniguchi}}, \bibinfo {author}
  {\bibfnamefont {K.}~\bibnamefont {Haule}}, \bibinfo {author} {\bibfnamefont
  {J.}~\bibnamefont {Mao}}, \ and\ \bibinfo {author} {\bibfnamefont {E.~Y.}\
  \bibnamefont {Andrei}},\ }\href {https://doi.org/10.1038/s41586-019-1460-4}
  {\bibfield  {journal} {\bibinfo  {journal} {Nature}\ } (\bibinfo {year}
  {2019})}\BibitemShut {NoStop}%
\bibitem [{\citenamefont {Shen}\ \emph {et~al.}()\citenamefont {Shen},
  \citenamefont {Li}, \citenamefont {Wang}, \citenamefont {Zhao}, \citenamefont
  {Tang}, \citenamefont {Liu}, \citenamefont {Tian}, \citenamefont {Chu},
  \citenamefont {Watanabe}, \citenamefont {Taniguchi} \emph
  {et~al.}}]{shen2019observation}%
  \BibitemOpen
  \bibfield  {author} {\bibinfo {author} {\bibfnamefont {C.}~\bibnamefont
  {Shen}}, \bibinfo {author} {\bibfnamefont {N.}~\bibnamefont {Li}}, \bibinfo
  {author} {\bibfnamefont {S.}~\bibnamefont {Wang}}, \bibinfo {author}
  {\bibfnamefont {Y.}~\bibnamefont {Zhao}}, \bibinfo {author} {\bibfnamefont
  {J.}~\bibnamefont {Tang}}, \bibinfo {author} {\bibfnamefont {J.}~\bibnamefont
  {Liu}}, \bibinfo {author} {\bibfnamefont {J.}~\bibnamefont {Tian}}, \bibinfo
  {author} {\bibfnamefont {Y.}~\bibnamefont {Chu}}, \bibinfo {author}
  {\bibfnamefont {K.}~\bibnamefont {Watanabe}}, \bibinfo {author}
  {\bibfnamefont {T.}~\bibnamefont {Taniguchi}},  \emph {et~al.},\ }\href
  {https://arxiv.org/abs/1903.06952} {\bibinfo  {journal} {arXiv:1903.06952}\
  }\BibitemShut {NoStop}%
\bibitem [{\citenamefont {Liu}\ \emph {et~al.}()\citenamefont {Liu},
  \citenamefont {Hao}, \citenamefont {Khalaf}, \citenamefont {Lee},
  \citenamefont {Watanabe}, \citenamefont {Taniguchi}, \citenamefont
  {Vishwanath},\ and\ \citenamefont {Kim}}]{liu2019spin}%
  \BibitemOpen
\bibfield  {journal} {  }\bibfield  {author} {\bibinfo {author} {\bibfnamefont
  {X.}~\bibnamefont {Liu}}, \bibinfo {author} {\bibfnamefont {Z.}~\bibnamefont
  {Hao}}, \bibinfo {author} {\bibfnamefont {E.}~\bibnamefont {Khalaf}},
  \bibinfo {author} {\bibfnamefont {J.~Y.}\ \bibnamefont {Lee}}, \bibinfo
  {author} {\bibfnamefont {K.}~\bibnamefont {Watanabe}}, \bibinfo {author}
  {\bibfnamefont {T.}~\bibnamefont {Taniguchi}}, \bibinfo {author}
  {\bibfnamefont {A.}~\bibnamefont {Vishwanath}}, \ and\ \bibinfo {author}
  {\bibfnamefont {P.}~\bibnamefont {Kim}},\ }\href
  {https://arxiv.org/abs/1903.08130} {\bibinfo  {journal} {arXiv:1903.08130}\
  }\BibitemShut {NoStop}%
\bibitem [{\citenamefont {Cao}\ \emph {et~al.}({\natexlab{b}})\citenamefont
  {Cao}, \citenamefont {Rodan-Legrain}, \citenamefont {Rubies-Bigord{\`a}},
  \citenamefont {Park}, \citenamefont {Watanabe}, \citenamefont {Taniguchi},\
  and\ \citenamefont {Jarillo-Herrero}}]{cao2019electric}%
  \BibitemOpen
\bibfield  {journal} {  }\bibfield  {author} {\bibinfo {author} {\bibfnamefont
  {Y.}~\bibnamefont {Cao}}, \bibinfo {author} {\bibfnamefont {D.}~\bibnamefont
  {Rodan-Legrain}}, \bibinfo {author} {\bibfnamefont {O.}~\bibnamefont
  {Rubies-Bigord{\`a}}}, \bibinfo {author} {\bibfnamefont {J.~M.}\ \bibnamefont
  {Park}}, \bibinfo {author} {\bibfnamefont {K.}~\bibnamefont {Watanabe}},
  \bibinfo {author} {\bibfnamefont {T.}~\bibnamefont {Taniguchi}}, \ and\
  \bibinfo {author} {\bibfnamefont {P.}~\bibnamefont {Jarillo-Herrero}},\
  }\href {https://arxiv.org/abs/1903.08596} {\bibfield  {journal} {\bibinfo
  {journal} {arXiv:1903.08596}\ } ({\natexlab{b}})}\BibitemShut {NoStop}%
\bibitem [{\citenamefont {Xu}\ and\ \citenamefont
  {Balents}(2018)}]{Balents2018}%
  \BibitemOpen
  \bibfield  {author} {\bibinfo {author} {\bibfnamefont {C.}~\bibnamefont
  {Xu}}\ and\ \bibinfo {author} {\bibfnamefont {L.}~\bibnamefont {Balents}},\
  }\href {\doibase 10.1103/PhysRevLett.121.087001} {\bibfield  {journal}
  {\bibinfo  {journal} {Phys. Rev. Lett.}\ }\textbf {\bibinfo {volume} {121}},\
  \bibinfo {pages} {087001} (\bibinfo {year} {2018})}\BibitemShut {NoStop}%
\bibitem [{\citenamefont {Po}\ \emph {et~al.}(2018)\citenamefont {Po},
  \citenamefont {Zou}, \citenamefont {Vishwanath},\ and\ \citenamefont
  {Senthil}}]{Senthil2018}%
  \BibitemOpen
  \bibfield  {author} {\bibinfo {author} {\bibfnamefont {H.~C.}\ \bibnamefont
  {Po}}, \bibinfo {author} {\bibfnamefont {L.}~\bibnamefont {Zou}}, \bibinfo
  {author} {\bibfnamefont {A.}~\bibnamefont {Vishwanath}}, \ and\ \bibinfo
  {author} {\bibfnamefont {T.}~\bibnamefont {Senthil}},\ }\href {\doibase
  10.1103/PhysRevX.8.031089} {\bibfield  {journal} {\bibinfo  {journal} {Phys.
  Rev. X}\ }\textbf {\bibinfo {volume} {8}},\ \bibinfo {pages} {031089}
  (\bibinfo {year} {2018})}\BibitemShut {NoStop}%
\bibitem [{\citenamefont {Koshino}\ \emph {et~al.}(2018)\citenamefont
  {Koshino}, \citenamefont {Yuan}, \citenamefont {Koretsune}, \citenamefont
  {Ochi}, \citenamefont {Kuroki},\ and\ \citenamefont {Fu}}]{Koshino2018}%
  \BibitemOpen
  \bibfield  {author} {\bibinfo {author} {\bibfnamefont {M.}~\bibnamefont
  {Koshino}}, \bibinfo {author} {\bibfnamefont {N.~F.~Q.}\ \bibnamefont
  {Yuan}}, \bibinfo {author} {\bibfnamefont {T.}~\bibnamefont {Koretsune}},
  \bibinfo {author} {\bibfnamefont {M.}~\bibnamefont {Ochi}}, \bibinfo {author}
  {\bibfnamefont {K.}~\bibnamefont {Kuroki}}, \ and\ \bibinfo {author}
  {\bibfnamefont {L.}~\bibnamefont {Fu}},\ }\href {\doibase
  10.1103/PhysRevX.8.031087} {\bibfield  {journal} {\bibinfo  {journal} {Phys.
  Rev. X}\ }\textbf {\bibinfo {volume} {8}},\ \bibinfo {pages} {031087}
  (\bibinfo {year} {2018})}\BibitemShut {NoStop}%
\bibitem [{\citenamefont {Kang}\ and\ \citenamefont {Vafek}(2018)}]{Kang2018}%
  \BibitemOpen
  \bibfield  {author} {\bibinfo {author} {\bibfnamefont {J.}~\bibnamefont
  {Kang}}\ and\ \bibinfo {author} {\bibfnamefont {O.}~\bibnamefont {Vafek}},\
  }\href {\doibase 10.1103/PhysRevX.8.031088} {\bibfield  {journal} {\bibinfo
  {journal} {Phys. Rev. X}\ }\textbf {\bibinfo {volume} {8}},\ \bibinfo {pages}
  {031088} (\bibinfo {year} {2018})}\BibitemShut {NoStop}%
\bibitem [{\citenamefont {Liu}\ \emph {et~al.}(2018)\citenamefont {Liu},
  \citenamefont {Zhang}, \citenamefont {Chen},\ and\ \citenamefont
  {Yang}}]{Liu2018chiral}%
  \BibitemOpen
  \bibfield  {author} {\bibinfo {author} {\bibfnamefont {C.-C.}\ \bibnamefont
  {Liu}}, \bibinfo {author} {\bibfnamefont {L.-D.}\ \bibnamefont {Zhang}},
  \bibinfo {author} {\bibfnamefont {W.-Q.}\ \bibnamefont {Chen}}, \ and\
  \bibinfo {author} {\bibfnamefont {F.}~\bibnamefont {Yang}},\ }\href {\doibase
  10.1103/PhysRevLett.121.217001} {\bibfield  {journal} {\bibinfo  {journal}
  {Phys. Rev. Lett.}\ }\textbf {\bibinfo {volume} {121}},\ \bibinfo {pages}
  {217001} (\bibinfo {year} {2018})}\BibitemShut {NoStop}%
\bibitem [{\citenamefont {Dodaro}\ \emph {et~al.}(2018)\citenamefont {Dodaro},
  \citenamefont {Kivelson}, \citenamefont {Schattner}, \citenamefont {Sun},\
  and\ \citenamefont {Wang}}]{Dodaro12018}%
  \BibitemOpen
  \bibfield  {author} {\bibinfo {author} {\bibfnamefont {J.~F.}\ \bibnamefont
  {Dodaro}}, \bibinfo {author} {\bibfnamefont {S.~A.}\ \bibnamefont
  {Kivelson}}, \bibinfo {author} {\bibfnamefont {Y.}~\bibnamefont {Schattner}},
  \bibinfo {author} {\bibfnamefont {X.~Q.}\ \bibnamefont {Sun}}, \ and\
  \bibinfo {author} {\bibfnamefont {C.}~\bibnamefont {Wang}},\ }\href {\doibase
  10.1103/PhysRevB.98.075154} {\bibfield  {journal} {\bibinfo  {journal} {Phys.
  Rev. B}\ }\textbf {\bibinfo {volume} {98}},\ \bibinfo {pages} {075154}
  (\bibinfo {year} {2018})}\BibitemShut {NoStop}%
\bibitem [{\citenamefont {Isobe}\ \emph {et~al.}(2018)\citenamefont {Isobe},
  \citenamefont {Yuan},\ and\ \citenamefont {Fu}}]{Isobe2018}%
  \BibitemOpen
  \bibfield  {author} {\bibinfo {author} {\bibfnamefont {H.}~\bibnamefont
  {Isobe}}, \bibinfo {author} {\bibfnamefont {N.~F.~Q.}\ \bibnamefont {Yuan}},
  \ and\ \bibinfo {author} {\bibfnamefont {L.}~\bibnamefont {Fu}},\ }\href
  {\doibase 10.1103/PhysRevX.8.041041} {\bibfield  {journal} {\bibinfo
  {journal} {Phys. Rev. X}\ }\textbf {\bibinfo {volume} {8}},\ \bibinfo {pages}
  {041041} (\bibinfo {year} {2018})}\BibitemShut {NoStop}%
\bibitem [{\citenamefont {You}\ and\ \citenamefont
  {Vishwanath}(2019)}]{You2018}%
  \BibitemOpen
  \bibfield  {author} {\bibinfo {author} {\bibfnamefont {Y.-Z.}\ \bibnamefont
  {You}}\ and\ \bibinfo {author} {\bibfnamefont {A.}~\bibnamefont
  {Vishwanath}},\ }\href {https://www.nature.com/articles/s41535-019-0153-4}
  {\bibfield  {journal} {\bibinfo  {journal} {npj Quantum Materials}\ }\textbf
  {\bibinfo {volume} {4}},\ \bibinfo {pages} {16} (\bibinfo {year}
  {2019})}\BibitemShut {NoStop}%
\bibitem [{\citenamefont {Tang}\ \emph {et~al.}(2019)\citenamefont {Tang},
  \citenamefont {Yang}, \citenamefont {Wang}, \citenamefont {Zhang},\ and\
  \citenamefont {Wang}}]{Tang2019}%
  \BibitemOpen
  \bibfield  {author} {\bibinfo {author} {\bibfnamefont {Q.-K.}\ \bibnamefont
  {Tang}}, \bibinfo {author} {\bibfnamefont {L.}~\bibnamefont {Yang}}, \bibinfo
  {author} {\bibfnamefont {D.}~\bibnamefont {Wang}}, \bibinfo {author}
  {\bibfnamefont {F.-C.}\ \bibnamefont {Zhang}}, \ and\ \bibinfo {author}
  {\bibfnamefont {Q.-H.}\ \bibnamefont {Wang}},\ }\href {\doibase
  10.1103/PhysRevB.99.094521} {\bibfield  {journal} {\bibinfo  {journal} {Phys.
  Rev. B}\ }\textbf {\bibinfo {volume} {99}},\ \bibinfo {pages} {094521}
  (\bibinfo {year} {2019})}\BibitemShut {NoStop}%
\bibitem [{\citenamefont {Rademaker}\ and\ \citenamefont
  {Mellado}(2018)}]{rademaker2018charge}%
  \BibitemOpen
  \bibfield  {author} {\bibinfo {author} {\bibfnamefont {L.}~\bibnamefont
  {Rademaker}}\ and\ \bibinfo {author} {\bibfnamefont {P.}~\bibnamefont
  {Mellado}},\ }\href {\doibase 10.1103/PhysRevB.98.235158} {\bibfield
  {journal} {\bibinfo  {journal} {Phys. Rev. B}\ }\textbf {\bibinfo {volume}
  {98}},\ \bibinfo {pages} {235158} (\bibinfo {year} {2018})}\BibitemShut
  {NoStop}%
\bibitem [{\citenamefont {Guinea}\ and\ \citenamefont
  {Walet}(2018)}]{guinea2018electrostatic}%
  \BibitemOpen
  \bibfield  {author} {\bibinfo {author} {\bibfnamefont {F.}~\bibnamefont
  {Guinea}}\ and\ \bibinfo {author} {\bibfnamefont {N.~R.}\ \bibnamefont
  {Walet}},\ }\href {https://www.pnas.org/content/115/52/13174} {\bibfield
  {journal} {\bibinfo  {journal} {Proc. Natl. Acad. Sci. U.S.A.}\ }\textbf
  {\bibinfo {volume} {115}},\ \bibinfo {pages} {13174} (\bibinfo {year}
  {2018})}\BibitemShut {NoStop}%
\bibitem [{\citenamefont {Gonz\'alez}\ and\ \citenamefont
  {Stauber}(2019)}]{gonzalez2018kohn}%
  \BibitemOpen
  \bibfield  {author} {\bibinfo {author} {\bibfnamefont {J.}~\bibnamefont
  {Gonz\'alez}}\ and\ \bibinfo {author} {\bibfnamefont {T.}~\bibnamefont
  {Stauber}},\ }\href {\doibase 10.1103/PhysRevLett.122.026801} {\bibfield
  {journal} {\bibinfo  {journal} {Phys. Rev. Lett.}\ }\textbf {\bibinfo
  {volume} {122}},\ \bibinfo {pages} {026801} (\bibinfo {year}
  {2019})}\BibitemShut {NoStop}%
\bibitem [{\citenamefont {Su}\ and\ \citenamefont {Lin}(2018)}]{Lin2018}%
  \BibitemOpen
  \bibfield  {author} {\bibinfo {author} {\bibfnamefont {Y.}~\bibnamefont
  {Su}}\ and\ \bibinfo {author} {\bibfnamefont {S.-Z.}\ \bibnamefont {Lin}},\
  }\href {\doibase 10.1103/PhysRevB.98.195101} {\bibfield  {journal} {\bibinfo
  {journal} {Phys. Rev. B}\ }\textbf {\bibinfo {volume} {98}},\ \bibinfo
  {pages} {195101} (\bibinfo {year} {2018})}\BibitemShut {NoStop}%
\bibitem [{\citenamefont {Ramires}\ and\ \citenamefont
  {Lado}(2018)}]{Lado2018}%
  \BibitemOpen
  \bibfield  {author} {\bibinfo {author} {\bibfnamefont {A.}~\bibnamefont
  {Ramires}}\ and\ \bibinfo {author} {\bibfnamefont {J.~L.}\ \bibnamefont
  {Lado}},\ }\href {\doibase 10.1103/PhysRevLett.121.146801} {\bibfield
  {journal} {\bibinfo  {journal} {Phys. Rev. Lett.}\ }\textbf {\bibinfo
  {volume} {121}},\ \bibinfo {pages} {146801} (\bibinfo {year}
  {2018})}\BibitemShut {NoStop}%
\bibitem [{\citenamefont {Tarnopolsky}\ \emph {et~al.}(2019)\citenamefont
  {Tarnopolsky}, \citenamefont {Kruchkov},\ and\ \citenamefont
  {Vishwanath}}]{Vishwanath2018origin}%
  \BibitemOpen
  \bibfield  {author} {\bibinfo {author} {\bibfnamefont {G.}~\bibnamefont
  {Tarnopolsky}}, \bibinfo {author} {\bibfnamefont {A.~J.}\ \bibnamefont
  {Kruchkov}}, \ and\ \bibinfo {author} {\bibfnamefont {A.}~\bibnamefont
  {Vishwanath}},\ }\href {\doibase 10.1103/PhysRevLett.122.106405} {\bibfield
  {journal} {\bibinfo  {journal} {Phys. Rev. Lett.}\ }\textbf {\bibinfo
  {volume} {122}},\ \bibinfo {pages} {106405} (\bibinfo {year}
  {2019})}\BibitemShut {NoStop}%
\bibitem [{\citenamefont {Ahn}\ \emph {et~al.}(2019)\citenamefont {Ahn},
  \citenamefont {Park},\ and\ \citenamefont {Yang}}]{Ahn2018failure}%
  \BibitemOpen
  \bibfield  {author} {\bibinfo {author} {\bibfnamefont {J.}~\bibnamefont
  {Ahn}}, \bibinfo {author} {\bibfnamefont {S.}~\bibnamefont {Park}}, \ and\
  \bibinfo {author} {\bibfnamefont {B.-J.}\ \bibnamefont {Yang}},\ }\href
  {\doibase 10.1103/PhysRevX.9.021013} {\bibfield  {journal} {\bibinfo
  {journal} {Phys. Rev. X}\ }\textbf {\bibinfo {volume} {9}},\ \bibinfo {pages}
  {021013} (\bibinfo {year} {2019})}\BibitemShut {NoStop}%
\bibitem [{\citenamefont {Song}\ \emph {et~al.}(2019)\citenamefont {Song},
  \citenamefont {Wang}, \citenamefont {Shi}, \citenamefont {Li}, \citenamefont
  {Fang},\ and\ \citenamefont {Bernevig}}]{Bernevig2018Topology}%
  \BibitemOpen
  \bibfield  {author} {\bibinfo {author} {\bibfnamefont {Z.}~\bibnamefont
  {Song}}, \bibinfo {author} {\bibfnamefont {Z.}~\bibnamefont {Wang}}, \bibinfo
  {author} {\bibfnamefont {W.}~\bibnamefont {Shi}}, \bibinfo {author}
  {\bibfnamefont {G.}~\bibnamefont {Li}}, \bibinfo {author} {\bibfnamefont
  {C.}~\bibnamefont {Fang}}, \ and\ \bibinfo {author} {\bibfnamefont {B.~A.}\
  \bibnamefont {Bernevig}},\ }\href
  {https://link.aps.org/doi/10.1103/PhysRevLett.123.036401} {\bibfield
  {journal} {\bibinfo  {journal} {Phys. Rev. Lett.}\ }\textbf {\bibinfo
  {volume} {123}},\ \bibinfo {pages} {036401} (\bibinfo {year}
  {2019})}\BibitemShut {NoStop}%
\bibitem [{\citenamefont {Hejazi}\ \emph {et~al.}(2019)\citenamefont {Hejazi},
  \citenamefont {Liu}, \citenamefont {Shapourian}, \citenamefont {Chen},\ and\
  \citenamefont {Balents}}]{hejazi2018multiple}%
  \BibitemOpen
  \bibfield  {author} {\bibinfo {author} {\bibfnamefont {K.}~\bibnamefont
  {Hejazi}}, \bibinfo {author} {\bibfnamefont {C.}~\bibnamefont {Liu}},
  \bibinfo {author} {\bibfnamefont {H.}~\bibnamefont {Shapourian}}, \bibinfo
  {author} {\bibfnamefont {X.}~\bibnamefont {Chen}}, \ and\ \bibinfo {author}
  {\bibfnamefont {L.}~\bibnamefont {Balents}},\ }\href {\doibase
  10.1103/PhysRevB.99.035111} {\bibfield  {journal} {\bibinfo  {journal} {Phys.
  Rev. B}\ }\textbf {\bibinfo {volume} {99}},\ \bibinfo {pages} {035111}
  (\bibinfo {year} {2019})}\BibitemShut {NoStop}%
\bibitem [{\citenamefont {Sherkunov}\ and\ \citenamefont
  {Betouras}(2018)}]{sherkunov2018novel}%
  \BibitemOpen
  \bibfield  {author} {\bibinfo {author} {\bibfnamefont {Y.}~\bibnamefont
  {Sherkunov}}\ and\ \bibinfo {author} {\bibfnamefont {J.~J.}\ \bibnamefont
  {Betouras}},\ }\href {\doibase 10.1103/PhysRevB.98.205151} {\bibfield
  {journal} {\bibinfo  {journal} {Phys. Rev. B}\ }\textbf {\bibinfo {volume}
  {98}},\ \bibinfo {pages} {205151} (\bibinfo {year} {2018})}\BibitemShut
  {NoStop}%
\bibitem [{\citenamefont {Kang}\ and\ \citenamefont
  {Vafek}(2019)}]{kang2018strong}%
  \BibitemOpen
  \bibfield  {author} {\bibinfo {author} {\bibfnamefont {J.}~\bibnamefont
  {Kang}}\ and\ \bibinfo {author} {\bibfnamefont {O.}~\bibnamefont {Vafek}},\
  }\href {\doibase 10.1103/PhysRevLett.122.246401} {\bibfield  {journal}
  {\bibinfo  {journal} {Phys. Rev. Lett.}\ }\textbf {\bibinfo {volume} {122}},\
  \bibinfo {pages} {246401} (\bibinfo {year} {2019})}\BibitemShut {NoStop}%
\bibitem [{\citenamefont {Seo}\ \emph {et~al.}(2019)\citenamefont {Seo},
  \citenamefont {Kotov},\ and\ \citenamefont {Uchoa}}]{Seo2019}%
  \BibitemOpen
  \bibfield  {author} {\bibinfo {author} {\bibfnamefont {K.}~\bibnamefont
  {Seo}}, \bibinfo {author} {\bibfnamefont {V.~N.}\ \bibnamefont {Kotov}}, \
  and\ \bibinfo {author} {\bibfnamefont {B.}~\bibnamefont {Uchoa}},\ }\href
  {\doibase 10.1103/PhysRevLett.122.246402} {\bibfield  {journal} {\bibinfo
  {journal} {Phys. Rev. Lett.}\ }\textbf {\bibinfo {volume} {122}},\ \bibinfo
  {pages} {246402} (\bibinfo {year} {2019})}\BibitemShut {NoStop}%
\bibitem [{\citenamefont {Lin}\ and\ \citenamefont
  {Nandkishore}()}]{lin2019chiral}%
  \BibitemOpen
  \bibfield  {author} {\bibinfo {author} {\bibfnamefont {Y.-P.}\ \bibnamefont
  {Lin}}\ and\ \bibinfo {author} {\bibfnamefont {R.~M.}\ \bibnamefont
  {Nandkishore}},\ }\href {https://arxiv.org/abs/1901.00500} {\bibinfo
  {journal} {arXiv:1901.00500}\ }\BibitemShut {NoStop}%
\bibitem [{\citenamefont {Peltonen}\ \emph {et~al.}(2018)\citenamefont
  {Peltonen}, \citenamefont {Ojaj\"arvi},\ and\ \citenamefont
  {Heikkil\"a}}]{Heikkila2018}%
  \BibitemOpen
\bibfield  {journal} {  }\bibfield  {author} {\bibinfo {author} {\bibfnamefont
  {T.~J.}\ \bibnamefont {Peltonen}}, \bibinfo {author} {\bibfnamefont
  {R.}~\bibnamefont {Ojaj\"arvi}}, \ and\ \bibinfo {author} {\bibfnamefont
  {T.~T.}\ \bibnamefont {Heikkil\"a}},\ }\href {\doibase
  10.1103/PhysRevB.98.220504} {\bibfield  {journal} {\bibinfo  {journal} {Phys.
  Rev. B}\ }\textbf {\bibinfo {volume} {98}},\ \bibinfo {pages} {220504(R)}
  (\bibinfo {year} {2018})}\BibitemShut {NoStop}%
\bibitem [{\citenamefont {Lian}\ \emph {et~al.}(2019)\citenamefont {Lian},
  \citenamefont {Wang},\ and\ \citenamefont {Bernevig}}]{Lian2018twisted}%
  \BibitemOpen
  \bibfield  {author} {\bibinfo {author} {\bibfnamefont {B.}~\bibnamefont
  {Lian}}, \bibinfo {author} {\bibfnamefont {Z.}~\bibnamefont {Wang}}, \ and\
  \bibinfo {author} {\bibfnamefont {B.~A.}\ \bibnamefont {Bernevig}},\ }\href
  {https://link.aps.org/doi/10.1103/PhysRevLett.122.257002} {\bibfield
  {journal} {\bibinfo  {journal} {Phys. Rev. Lett.}\ }\textbf {\bibinfo
  {volume} {122}},\ \bibinfo {pages} {257002} (\bibinfo {year}
  {2019})}\BibitemShut {NoStop}%
\bibitem [{\citenamefont {Choi}\ and\ \citenamefont
  {Choi}(2018)}]{choi2018electron}%
  \BibitemOpen
  \bibfield  {author} {\bibinfo {author} {\bibfnamefont {Y.~W.}\ \bibnamefont
  {Choi}}\ and\ \bibinfo {author} {\bibfnamefont {H.~J.}\ \bibnamefont
  {Choi}},\ }\href {\doibase 10.1103/PhysRevB.98.241412} {\bibfield  {journal}
  {\bibinfo  {journal} {Phys. Rev. B}\ }\textbf {\bibinfo {volume} {98}},\
  \bibinfo {pages} {241412(R)} (\bibinfo {year} {2018})}\BibitemShut {NoStop}%
\bibitem [{\citenamefont {Wu}\ \emph {et~al.}(2018)\citenamefont {Wu},
  \citenamefont {MacDonald},\ and\ \citenamefont {Martin}}]{Wu2018phonon}%
  \BibitemOpen
  \bibfield  {author} {\bibinfo {author} {\bibfnamefont {F.}~\bibnamefont
  {Wu}}, \bibinfo {author} {\bibfnamefont {A.~H.}\ \bibnamefont {MacDonald}}, \
  and\ \bibinfo {author} {\bibfnamefont {I.}~\bibnamefont {Martin}},\ }\href
  {\doibase 10.1103/PhysRevLett.121.257001} {\bibfield  {journal} {\bibinfo
  {journal} {Phys. Rev. Lett.}\ }\textbf {\bibinfo {volume} {121}},\ \bibinfo
  {pages} {257001} (\bibinfo {year} {2018})}\BibitemShut {NoStop}%
\bibitem [{\citenamefont {Wu}\ \emph {et~al.}(2019{\natexlab{a}})\citenamefont
  {Wu}, \citenamefont {Hwang},\ and\ \citenamefont {Das~Sarma}}]{wu2019phonon}%
  \BibitemOpen
  \bibfield  {author} {\bibinfo {author} {\bibfnamefont {F.}~\bibnamefont
  {Wu}}, \bibinfo {author} {\bibfnamefont {E.}~\bibnamefont {Hwang}}, \ and\
  \bibinfo {author} {\bibfnamefont {S.}~\bibnamefont {Das~Sarma}},\ }\href
  {\doibase 10.1103/PhysRevB.99.165112} {\bibfield  {journal} {\bibinfo
  {journal} {Phys. Rev. B}\ }\textbf {\bibinfo {volume} {99}},\ \bibinfo
  {pages} {165112} (\bibinfo {year} {2019}{\natexlab{a}})}\BibitemShut
  {NoStop}%
\bibitem [{\citenamefont {Wu}(2019)}]{wu2018topological}%
  \BibitemOpen
  \bibfield  {author} {\bibinfo {author} {\bibfnamefont {F.}~\bibnamefont
  {Wu}},\ }\href {\doibase 10.1103/PhysRevB.99.195114} {\bibfield  {journal}
  {\bibinfo  {journal} {Phys. Rev. B}\ }\textbf {\bibinfo {volume} {99}},\
  \bibinfo {pages} {195114} (\bibinfo {year} {2019})}\BibitemShut {NoStop}%
\bibitem [{\citenamefont {Wu}\ and\ \citenamefont
  {Das~Sarma}(2019)}]{wu2019identification}%
  \BibitemOpen
  \bibfield  {author} {\bibinfo {author} {\bibfnamefont {F.}~\bibnamefont
  {Wu}}\ and\ \bibinfo {author} {\bibfnamefont {S.}~\bibnamefont {Das~Sarma}},\
  }\href {\doibase 10.1103/PhysRevB.99.220507} {\bibfield  {journal} {\bibinfo
  {journal} {Phys. Rev. B}\ }\textbf {\bibinfo {volume} {99}},\ \bibinfo
  {pages} {220507(R)} (\bibinfo {year} {2019})}\BibitemShut {NoStop}%
\bibitem [{\citenamefont {Wu}\ and\ \citenamefont {Das~Sarma}()}]{wu2019Ferro}%
  \BibitemOpen
  \bibfield  {author} {\bibinfo {author} {\bibfnamefont {F.}~\bibnamefont
  {Wu}}\ and\ \bibinfo {author} {\bibfnamefont {S.}~\bibnamefont {Das~Sarma}},\
  }\href {https://arxiv.org/abs/1906.07302} {\bibinfo  {journal}
  {arXiv:1906.07302}\ }\BibitemShut {NoStop}%
\bibitem [{\citenamefont {Zhang}\ \emph
  {et~al.}(2019{\natexlab{a}})\citenamefont {Zhang}, \citenamefont {Mao},
  \citenamefont {Cao}, \citenamefont {Jarillo-Herrero},\ and\ \citenamefont
  {Senthil}}]{Zhang2019}%
  \BibitemOpen
\bibfield  {journal} {  }\bibfield  {author} {\bibinfo {author} {\bibfnamefont
  {Y.-H.}\ \bibnamefont {Zhang}}, \bibinfo {author} {\bibfnamefont
  {D.}~\bibnamefont {Mao}}, \bibinfo {author} {\bibfnamefont {Y.}~\bibnamefont
  {Cao}}, \bibinfo {author} {\bibfnamefont {P.}~\bibnamefont
  {Jarillo-Herrero}}, \ and\ \bibinfo {author} {\bibfnamefont {T.}~\bibnamefont
  {Senthil}},\ }\href {\doibase 10.1103/PhysRevB.99.075127} {\bibfield
  {journal} {\bibinfo  {journal} {Phys. Rev. B}\ }\textbf {\bibinfo {volume}
  {99}},\ \bibinfo {pages} {075127} (\bibinfo {year}
  {2019}{\natexlab{a}})}\BibitemShut {NoStop}%
\bibitem [{\citenamefont {Chittari}\ \emph {et~al.}(2019)\citenamefont
  {Chittari}, \citenamefont {Chen}, \citenamefont {Zhang}, \citenamefont
  {Wang},\ and\ \citenamefont {Jung}}]{JungTopology}%
  \BibitemOpen
  \bibfield  {author} {\bibinfo {author} {\bibfnamefont {B.~L.}\ \bibnamefont
  {Chittari}}, \bibinfo {author} {\bibfnamefont {G.}~\bibnamefont {Chen}},
  \bibinfo {author} {\bibfnamefont {Y.}~\bibnamefont {Zhang}}, \bibinfo
  {author} {\bibfnamefont {F.}~\bibnamefont {Wang}}, \ and\ \bibinfo {author}
  {\bibfnamefont {J.}~\bibnamefont {Jung}},\ }\href {\doibase
  10.1103/PhysRevLett.122.016401} {\bibfield  {journal} {\bibinfo  {journal}
  {Phys. Rev. Lett.}\ }\textbf {\bibinfo {volume} {122}},\ \bibinfo {pages}
  {016401} (\bibinfo {year} {2019})}\BibitemShut {NoStop}%
\bibitem [{\citenamefont {Wu}\ \emph {et~al.}(2019{\natexlab{b}})\citenamefont
  {Wu}, \citenamefont {Lovorn}, \citenamefont {Tutuc}, \citenamefont {Martin},\
  and\ \citenamefont {MacDonald}}]{WuTITMD}%
  \BibitemOpen
  \bibfield  {author} {\bibinfo {author} {\bibfnamefont {F.}~\bibnamefont
  {Wu}}, \bibinfo {author} {\bibfnamefont {T.}~\bibnamefont {Lovorn}}, \bibinfo
  {author} {\bibfnamefont {E.}~\bibnamefont {Tutuc}}, \bibinfo {author}
  {\bibfnamefont {I.}~\bibnamefont {Martin}}, \ and\ \bibinfo {author}
  {\bibfnamefont {A.~H.}\ \bibnamefont {MacDonald}},\ }\href {\doibase
  10.1103/PhysRevLett.122.086402} {\bibfield  {journal} {\bibinfo  {journal}
  {Phys. Rev. Lett.}\ }\textbf {\bibinfo {volume} {122}},\ \bibinfo {pages}
  {086402} (\bibinfo {year} {2019}{\natexlab{b}})}\BibitemShut {NoStop}%
\bibitem [{\citenamefont {Xie}\ and\ \citenamefont
  {MacDonald}()}]{xie2018nature}%
  \BibitemOpen
  \bibfield  {author} {\bibinfo {author} {\bibfnamefont {M.}~\bibnamefont
  {Xie}}\ and\ \bibinfo {author} {\bibfnamefont {A.~H.}\ \bibnamefont
  {MacDonald}},\ }\href {https://arxiv.org/abs/1812.04213} {\bibinfo  {journal}
  {arXiv:1812.04213}\ }\BibitemShut {NoStop}%
\bibitem [{\citenamefont {Bultinck}\ \emph {et~al.}()\citenamefont {Bultinck},
  \citenamefont {Chatterjee},\ and\ \citenamefont
  {Zaletel}}]{bultinck2019anomalous}%
  \BibitemOpen
\bibfield  {journal} {  }\bibfield  {author} {\bibinfo {author} {\bibfnamefont
  {N.}~\bibnamefont {Bultinck}}, \bibinfo {author} {\bibfnamefont
  {S.}~\bibnamefont {Chatterjee}}, \ and\ \bibinfo {author} {\bibfnamefont
  {M.~P.}\ \bibnamefont {Zaletel}},\ }\href
  {https://arxiv.org/abs/1901.08110v2} {\bibinfo  {journal} {arXiv:1901.08110}\
  }\BibitemShut {NoStop}%
\bibitem [{\citenamefont {Zhang}\ \emph
  {et~al.}(2019{\natexlab{b}})\citenamefont {Zhang}, \citenamefont {Mao},\ and\
  \citenamefont {Senthil}}]{zhang2019twisted}%
  \BibitemOpen
\bibfield  {journal} {  }\bibfield  {author} {\bibinfo {author} {\bibfnamefont
  {Y.-H.}\ \bibnamefont {Zhang}}, \bibinfo {author} {\bibfnamefont
  {D.}~\bibnamefont {Mao}}, \ and\ \bibinfo {author} {\bibfnamefont
  {T.}~\bibnamefont {Senthil}},\ }\href {https://arxiv.org/abs/1901.08209}
  {\bibfield  {journal} {\bibinfo  {journal} {arXiv:1901.08209}\ } (\bibinfo
  {year} {2019}{\natexlab{b}})}\BibitemShut {NoStop}%
\bibitem [{\citenamefont {Liu}\ \emph {et~al.}(2019)\citenamefont {Liu},
  \citenamefont {Ma}, \citenamefont {Gao},\ and\ \citenamefont
  {Dai}}]{LiuMulti}%
  \BibitemOpen
  \bibfield  {author} {\bibinfo {author} {\bibfnamefont {J.}~\bibnamefont
  {Liu}}, \bibinfo {author} {\bibfnamefont {Z.}~\bibnamefont {Ma}}, \bibinfo
  {author} {\bibfnamefont {J.}~\bibnamefont {Gao}}, \ and\ \bibinfo {author}
  {\bibfnamefont {X.}~\bibnamefont {Dai}},\ }\href {\doibase
  10.1103/PhysRevX.9.031021} {\bibfield  {journal} {\bibinfo  {journal} {Phys.
  Rev. X}\ }\textbf {\bibinfo {volume} {9}},\ \bibinfo {pages} {031021}
  (\bibinfo {year} {2019})}\BibitemShut {NoStop}%
\bibitem [{\citenamefont {Lee}\ \emph {et~al.}()\citenamefont {Lee},
  \citenamefont {Khalaf}, \citenamefont {Liu}, \citenamefont {Liu},
  \citenamefont {Hao}, \citenamefont {Kim},\ and\ \citenamefont
  {Vishwanath}}]{lee2019theory}%
  \BibitemOpen
  \bibfield  {author} {\bibinfo {author} {\bibfnamefont {J.~Y.}\ \bibnamefont
  {Lee}}, \bibinfo {author} {\bibfnamefont {E.}~\bibnamefont {Khalaf}},
  \bibinfo {author} {\bibfnamefont {S.}~\bibnamefont {Liu}}, \bibinfo {author}
  {\bibfnamefont {X.}~\bibnamefont {Liu}}, \bibinfo {author} {\bibfnamefont
  {Z.}~\bibnamefont {Hao}}, \bibinfo {author} {\bibfnamefont {P.}~\bibnamefont
  {Kim}}, \ and\ \bibinfo {author} {\bibfnamefont {A.}~\bibnamefont
  {Vishwanath}},\ }\href {https://arxiv.org/abs/1903.08685} {\bibinfo
  {journal} {arXiv:1903.08685}\ }\BibitemShut {NoStop}%
\bibitem [{\citenamefont {Wu}\ \emph {et~al.}(2019{\natexlab{c}})\citenamefont
  {Wu}, \citenamefont {Keselman}, \citenamefont {Jian}, \citenamefont
  {Pawlak},\ and\ \citenamefont {Xu}}]{Xu2019Ferro}%
  \BibitemOpen
\bibfield  {journal} {  }\bibfield  {author} {\bibinfo {author} {\bibfnamefont
  {X.-C.}\ \bibnamefont {Wu}}, \bibinfo {author} {\bibfnamefont
  {A.}~\bibnamefont {Keselman}}, \bibinfo {author} {\bibfnamefont {C.-M.}\
  \bibnamefont {Jian}}, \bibinfo {author} {\bibfnamefont {K.~A.}\ \bibnamefont
  {Pawlak}}, \ and\ \bibinfo {author} {\bibfnamefont {C.}~\bibnamefont {Xu}},\
  }\href {\doibase 10.1103/PhysRevB.100.024421} {\bibfield  {journal} {\bibinfo
   {journal} {Phys. Rev. B}\ }\textbf {\bibinfo {volume} {100}},\ \bibinfo
  {pages} {024421} (\bibinfo {year} {2019}{\natexlab{c}})}\BibitemShut
  {NoStop}%
\bibitem [{\citenamefont {Hazra}\ \emph {et~al.}()\citenamefont {Hazra},
  \citenamefont {Verma},\ and\ \citenamefont {Randeria}}]{hazra2018upper}%
  \BibitemOpen
  \bibfield  {author} {\bibinfo {author} {\bibfnamefont {T.}~\bibnamefont
  {Hazra}}, \bibinfo {author} {\bibfnamefont {N.}~\bibnamefont {Verma}}, \ and\
  \bibinfo {author} {\bibfnamefont {M.}~\bibnamefont {Randeria}},\ }\href
  {https://arxiv.org/abs/1811.12428} {\bibinfo  {journal} {arXiv:1811.12428}\
  }\BibitemShut {NoStop}%
\bibitem [{\citenamefont {Xie}\ \emph {et~al.}()\citenamefont {Xie},
  \citenamefont {Song}, \citenamefont {Lian},\ and\ \citenamefont
  {Bernevig}}]{xie2019topology}%
  \BibitemOpen
\bibfield  {journal} {  }\bibfield  {author} {\bibinfo {author} {\bibfnamefont
  {F.}~\bibnamefont {Xie}}, \bibinfo {author} {\bibfnamefont {Z.}~\bibnamefont
  {Song}}, \bibinfo {author} {\bibfnamefont {B.}~\bibnamefont {Lian}}, \ and\
  \bibinfo {author} {\bibfnamefont {B.~A.}\ \bibnamefont {Bernevig}},\ }\href
  {https://arxiv.org/abs/1906.02213} {\bibinfo  {journal} {arXiv:1906.02213}\
  }\BibitemShut {NoStop}%
\bibitem [{\citenamefont {Julku}\ \emph {et~al.}()\citenamefont {Julku},
  \citenamefont {Peltonen}, \citenamefont {Liang}, \citenamefont
  {Heikkil{\"a}},\ and\ \citenamefont {T{\"o}rm{\"a}}}]{julku2019superfluid}%
  \BibitemOpen
\bibfield  {journal} {  }\bibfield  {author} {\bibinfo {author} {\bibfnamefont
  {A.}~\bibnamefont {Julku}}, \bibinfo {author} {\bibfnamefont {T.~J.}\
  \bibnamefont {Peltonen}}, \bibinfo {author} {\bibfnamefont {L.}~\bibnamefont
  {Liang}}, \bibinfo {author} {\bibfnamefont {T.~T.}\ \bibnamefont
  {Heikkil{\"a}}}, \ and\ \bibinfo {author} {\bibfnamefont {P.}~\bibnamefont
  {T{\"o}rm{\"a}}},\ }\href {https://arxiv.org/abs/1906.06313} {\bibinfo
  {journal} {arXiv:1906.06313}\ }\BibitemShut {NoStop}%
\bibitem [{\citenamefont {Hu}\ \emph {et~al.}()\citenamefont {Hu},
  \citenamefont {Hyart}, \citenamefont {Pikulin},\ and\ \citenamefont
  {Rossi}}]{hu2019geometric}%
  \BibitemOpen
\bibfield  {journal} {  }\bibfield  {author} {\bibinfo {author} {\bibfnamefont
  {X.}~\bibnamefont {Hu}}, \bibinfo {author} {\bibfnamefont {T.}~\bibnamefont
  {Hyart}}, \bibinfo {author} {\bibfnamefont {D.~I.}\ \bibnamefont {Pikulin}},
  \ and\ \bibinfo {author} {\bibfnamefont {E.}~\bibnamefont {Rossi}},\ }\href
  {https://arxiv.org/abs/1906.07152} {\bibinfo  {journal} {arXiv:1906.07152}\
  }\BibitemShut {NoStop}%
\bibitem [{\citenamefont {Liu}\ and\ \citenamefont {Dai}()}]{liu2019anomalous}%
  \BibitemOpen
\bibfield  {journal} {  }\bibfield  {author} {\bibinfo {author} {\bibfnamefont
  {J.}~\bibnamefont {Liu}}\ and\ \bibinfo {author} {\bibfnamefont
  {X.}~\bibnamefont {Dai}},\ }\href {https://arxiv.org/abs/1907.08932}
  {\bibinfo  {journal} {arXiv:1907.08932}\ }\BibitemShut {NoStop}%
\bibitem [{\citenamefont {Chang}\ \emph {et~al.}(2013)\citenamefont {Chang},
  \citenamefont {Zhang}, \citenamefont {Feng}, \citenamefont {Shen},
  \citenamefont {Zhang}, \citenamefont {Guo}, \citenamefont {Li}, \citenamefont
  {Ou}, \citenamefont {Wei}, \citenamefont {Wang}, \citenamefont {Ji},
  \citenamefont {Feng}, \citenamefont {Ji}, \citenamefont {Chen}, \citenamefont
  {Jia}, \citenamefont {Dai}, \citenamefont {Fang}, \citenamefont {Zhang},
  \citenamefont {He}, \citenamefont {Wang}, \citenamefont {Lu}, \citenamefont
  {Ma},\ and\ \citenamefont {Xue}}]{Chang167}%
  \BibitemOpen
\bibfield  {journal} {  }\bibfield  {author} {\bibinfo {author} {\bibfnamefont
  {C.-Z.}\ \bibnamefont {Chang}}, \bibinfo {author} {\bibfnamefont
  {J.}~\bibnamefont {Zhang}}, \bibinfo {author} {\bibfnamefont
  {X.}~\bibnamefont {Feng}}, \bibinfo {author} {\bibfnamefont {J.}~\bibnamefont
  {Shen}}, \bibinfo {author} {\bibfnamefont {Z.}~\bibnamefont {Zhang}},
  \bibinfo {author} {\bibfnamefont {M.}~\bibnamefont {Guo}}, \bibinfo {author}
  {\bibfnamefont {K.}~\bibnamefont {Li}}, \bibinfo {author} {\bibfnamefont
  {Y.}~\bibnamefont {Ou}}, \bibinfo {author} {\bibfnamefont {P.}~\bibnamefont
  {Wei}}, \bibinfo {author} {\bibfnamefont {L.-L.}\ \bibnamefont {Wang}},
  \bibinfo {author} {\bibfnamefont {Z.-Q.}\ \bibnamefont {Ji}}, \bibinfo
  {author} {\bibfnamefont {Y.}~\bibnamefont {Feng}}, \bibinfo {author}
  {\bibfnamefont {S.}~\bibnamefont {Ji}}, \bibinfo {author} {\bibfnamefont
  {X.}~\bibnamefont {Chen}}, \bibinfo {author} {\bibfnamefont {J.}~\bibnamefont
  {Jia}}, \bibinfo {author} {\bibfnamefont {X.}~\bibnamefont {Dai}}, \bibinfo
  {author} {\bibfnamefont {Z.}~\bibnamefont {Fang}}, \bibinfo {author}
  {\bibfnamefont {S.-C.}\ \bibnamefont {Zhang}}, \bibinfo {author}
  {\bibfnamefont {K.}~\bibnamefont {He}}, \bibinfo {author} {\bibfnamefont
  {Y.}~\bibnamefont {Wang}}, \bibinfo {author} {\bibfnamefont {L.}~\bibnamefont
  {Lu}}, \bibinfo {author} {\bibfnamefont {X.-C.}\ \bibnamefont {Ma}}, \ and\
  \bibinfo {author} {\bibfnamefont {Q.-K.}\ \bibnamefont {Xue}},\ }\href
  {\doibase 10.1126/science.1234414} {\bibfield  {journal} {\bibinfo  {journal}
  {Science}\ }\textbf {\bibinfo {volume} {340}},\ \bibinfo {pages} {167}
  (\bibinfo {year} {2013})}\BibitemShut {NoStop}%
\bibitem [{\citenamefont {Moon}\ \emph {et~al.}(1995)\citenamefont {Moon},
  \citenamefont {Mori}, \citenamefont {Yang}, \citenamefont {Girvin},
  \citenamefont {MacDonald}, \citenamefont {Zheng}, \citenamefont {Yoshioka},\
  and\ \citenamefont {Zhang}}]{Moon1995}%
  \BibitemOpen
  \bibfield  {author} {\bibinfo {author} {\bibfnamefont {K.}~\bibnamefont
  {Moon}}, \bibinfo {author} {\bibfnamefont {H.}~\bibnamefont {Mori}}, \bibinfo
  {author} {\bibfnamefont {K.}~\bibnamefont {Yang}}, \bibinfo {author}
  {\bibfnamefont {S.~M.}\ \bibnamefont {Girvin}}, \bibinfo {author}
  {\bibfnamefont {A.~H.}\ \bibnamefont {MacDonald}}, \bibinfo {author}
  {\bibfnamefont {L.}~\bibnamefont {Zheng}}, \bibinfo {author} {\bibfnamefont
  {D.}~\bibnamefont {Yoshioka}}, \ and\ \bibinfo {author} {\bibfnamefont
  {S.-C.}\ \bibnamefont {Zhang}},\ }\href {\doibase 10.1103/PhysRevB.51.5138}
  {\bibfield  {journal} {\bibinfo  {journal} {Phys. Rev. B}\ }\textbf {\bibinfo
  {volume} {51}},\ \bibinfo {pages} {5138} (\bibinfo {year}
  {1995})}\BibitemShut {NoStop}%
\bibitem [{\citenamefont {Yang}\ \emph {et~al.}(2006)\citenamefont {Yang},
  \citenamefont {Das~Sarma},\ and\ \citenamefont {MacDonald}}]{Yang2006}%
  \BibitemOpen
  \bibfield  {author} {\bibinfo {author} {\bibfnamefont {K.}~\bibnamefont
  {Yang}}, \bibinfo {author} {\bibfnamefont {S.}~\bibnamefont {Das~Sarma}}, \
  and\ \bibinfo {author} {\bibfnamefont {A.~H.}\ \bibnamefont {MacDonald}},\
  }\href {\doibase 10.1103/PhysRevB.74.075423} {\bibfield  {journal} {\bibinfo
  {journal} {Phys. Rev. B}\ }\textbf {\bibinfo {volume} {74}},\ \bibinfo
  {pages} {075423} (\bibinfo {year} {2006})}\BibitemShut {NoStop}%
\bibitem [{\citenamefont {Bistritzer}\ and\ \citenamefont
  {MacDonald}(2011)}]{Bistritzer2011}%
  \BibitemOpen
  \bibfield  {author} {\bibinfo {author} {\bibfnamefont {R.}~\bibnamefont
  {Bistritzer}}\ and\ \bibinfo {author} {\bibfnamefont {A.~H.}\ \bibnamefont
  {MacDonald}},\ }\href {http://www.pnas.org/content/108/30/12233.abstract}
  {\bibfield  {journal} {\bibinfo  {journal} {Proc. Natl. Acad. Sci. U.S.A.}\
  }\textbf {\bibinfo {volume} {108}},\ \bibinfo {pages} {12233} (\bibinfo
  {year} {2011})}\BibitemShut {NoStop}%
\bibitem [{SM()}]{SM}%
  \BibitemOpen
  \href@noop {} {\bibinfo  {journal} {See Supplemental Material at URL for
  details of the TBG moir\'e Hamiltonian and an analysis of QAHF in TBG that is
  closely aligned to both the top and bottom encapsulating hBN layers.}\
  }\BibitemShut {NoStop}%
\bibitem [{\citenamefont {Hunt}\ \emph {et~al.}(2013)\citenamefont {Hunt},
  \citenamefont {Sanchez-Yamagishi}, \citenamefont {Young}, \citenamefont
  {Yankowitz}, \citenamefont {LeRoy}, \citenamefont {Watanabe}, \citenamefont
  {Taniguchi}, \citenamefont {Moon}, \citenamefont {Koshino}, \citenamefont
  {Jarillo-Herrero},\ and\ \citenamefont {Ashoori}}]{Hunt2013}%
  \BibitemOpen
\bibfield  {journal} {  }\bibfield  {author} {\bibinfo {author} {\bibfnamefont
  {B.}~\bibnamefont {Hunt}}, \bibinfo {author} {\bibfnamefont {J.~D.}\
  \bibnamefont {Sanchez-Yamagishi}}, \bibinfo {author} {\bibfnamefont {A.~F.}\
  \bibnamefont {Young}}, \bibinfo {author} {\bibfnamefont {M.}~\bibnamefont
  {Yankowitz}}, \bibinfo {author} {\bibfnamefont {B.~J.}\ \bibnamefont
  {LeRoy}}, \bibinfo {author} {\bibfnamefont {K.}~\bibnamefont {Watanabe}},
  \bibinfo {author} {\bibfnamefont {T.}~\bibnamefont {Taniguchi}}, \bibinfo
  {author} {\bibfnamefont {P.}~\bibnamefont {Moon}}, \bibinfo {author}
  {\bibfnamefont {M.}~\bibnamefont {Koshino}}, \bibinfo {author} {\bibfnamefont
  {P.}~\bibnamefont {Jarillo-Herrero}}, \ and\ \bibinfo {author} {\bibfnamefont
  {R.~C.}\ \bibnamefont {Ashoori}},\ }\href {\doibase 10.1126/science.1237240}
  {\bibfield  {journal} {\bibinfo  {journal} {Science}\ }\textbf {\bibinfo
  {volume} {340}},\ \bibinfo {pages} {1427} (\bibinfo {year}
  {2013})}\BibitemShut {NoStop}%
\bibitem [{\citenamefont {Srivastava}\ and\ \citenamefont
  {Imamo\u{g}lu}(2015)}]{Srivastava2015}%
  \BibitemOpen
  \bibfield  {author} {\bibinfo {author} {\bibfnamefont {A.}~\bibnamefont
  {Srivastava}}\ and\ \bibinfo {author} {\bibfnamefont {A.}~\bibnamefont
  {Imamo\u{g}lu}},\ }\href {\doibase 10.1103/PhysRevLett.115.166802} {\bibfield
   {journal} {\bibinfo  {journal} {Phys. Rev. Lett.}\ }\textbf {\bibinfo
  {volume} {115}},\ \bibinfo {pages} {166802} (\bibinfo {year}
  {2015})}\BibitemShut {NoStop}%
\bibitem [{\citenamefont {Girvin}()}]{girvin1999quantum}%
  \BibitemOpen
  \bibfield  {author} {\bibinfo {author} {\bibfnamefont {S.~M.}\ \bibnamefont
  {Girvin}},\ }\href
  {https://link.springer.com/chapter/10.1007/3-540-46637-1_2} {\bibinfo
  {journal} {Topological aspects of low dimensional systems(Springer,
  1999),53--175}\ }\BibitemShut {NoStop}%
\bibitem [{\citenamefont {Chaikin}\ and\ \citenamefont {Lubensky}()}]{Chaiken}%
  \BibitemOpen
\bibfield  {journal} {  }\bibfield  {author} {\bibinfo {author} {\bibfnamefont
  {P.}~\bibnamefont {Chaikin}}\ and\ \bibinfo {author} {\bibfnamefont {T.~C.}\
  \bibnamefont {Lubensky}},\ }\href@noop {} {\bibinfo  {journal} {Principles of
  Condensed Matter Physics (Cambridge University Press, Cambridge, England,
  2000)}\ }\BibitemShut {NoStop}%
\bibitem [{\citenamefont {Li}\ \emph {et~al.}(2014)\citenamefont {Li},
  \citenamefont {Zhang}, \citenamefont {Niu},\ and\ \citenamefont
  {MacDonald}}]{XiaoLi2014}%
  \BibitemOpen
\bibfield  {journal} {  }\bibfield  {author} {\bibinfo {author} {\bibfnamefont
  {X.}~\bibnamefont {Li}}, \bibinfo {author} {\bibfnamefont {F.}~\bibnamefont
  {Zhang}}, \bibinfo {author} {\bibfnamefont {Q.}~\bibnamefont {Niu}}, \ and\
  \bibinfo {author} {\bibfnamefont {A.~H.}\ \bibnamefont {MacDonald}},\ }\href
  {\doibase 10.1103/PhysRevLett.113.116803} {\bibfield  {journal} {\bibinfo
  {journal} {Phys. Rev. Lett.}\ }\textbf {\bibinfo {volume} {113}},\ \bibinfo
  {pages} {116803} (\bibinfo {year} {2014})}\BibitemShut {NoStop}%
\bibitem [{\citenamefont {Kato}\ \emph {et~al.}(2010)\citenamefont {Kato},
  \citenamefont {Martin},\ and\ \citenamefont {Batista}}]{Kato2010}%
  \BibitemOpen
  \bibfield  {author} {\bibinfo {author} {\bibfnamefont {Y.}~\bibnamefont
  {Kato}}, \bibinfo {author} {\bibfnamefont {I.}~\bibnamefont {Martin}}, \ and\
  \bibinfo {author} {\bibfnamefont {C.~D.}\ \bibnamefont {Batista}},\ }\href
  {\doibase 10.1103/PhysRevLett.105.266405} {\bibfield  {journal} {\bibinfo
  {journal} {Phys. Rev. Lett.}\ }\textbf {\bibinfo {volume} {105}},\ \bibinfo
  {pages} {266405} (\bibinfo {year} {2010})}\BibitemShut {NoStop}%
\bibitem [{\citenamefont {Repellin}\ \emph {et~al.}()\citenamefont {Repellin},
  \citenamefont {Dong}, \citenamefont {Zhang},\ and\ \citenamefont
  {Senthil}}]{repellin2019ferromagnetism}%
  \BibitemOpen
  \bibfield  {author} {\bibinfo {author} {\bibfnamefont {C.}~\bibnamefont
  {Repellin}}, \bibinfo {author} {\bibfnamefont {Z.}~\bibnamefont {Dong}},
  \bibinfo {author} {\bibfnamefont {Y.-H.}\ \bibnamefont {Zhang}}, \ and\
  \bibinfo {author} {\bibfnamefont {T.}~\bibnamefont {Senthil}},\ }\href
  {https://arxiv.org/abs/1907.11723} {\bibinfo  {journal} {arXiv:1907.11723}\
  }\BibitemShut {NoStop}%
\bibitem [{\citenamefont {Alavirad}\ and\ \citenamefont
  {Sau}()}]{alavirad2019ferromagnetism}%
  \BibitemOpen
\bibfield  {journal} {  }\bibfield  {author} {\bibinfo {author} {\bibfnamefont
  {Y.}~\bibnamefont {Alavirad}}\ and\ \bibinfo {author} {\bibfnamefont {J.~D.}\
  \bibnamefont {Sau}},\ }\href {https://arxiv.org/abs/1907.13633} {\bibinfo
  {journal} {arXiv:1907.13633}\ }\BibitemShut {NoStop}%
\bibitem [{\citenamefont {Chatterjee}\ \emph {et~al.}()\citenamefont
  {Chatterjee}, \citenamefont {Bultinck},\ and\ \citenamefont
  {Zaletel}}]{chatterjee2019symmetry}%
  \BibitemOpen
\bibfield  {journal} {  }\bibfield  {author} {\bibinfo {author} {\bibfnamefont
  {S.}~\bibnamefont {Chatterjee}}, \bibinfo {author} {\bibfnamefont
  {N.}~\bibnamefont {Bultinck}}, \ and\ \bibinfo {author} {\bibfnamefont
  {M.~P.}\ \bibnamefont {Zaletel}},\ }\href {https://arxiv.org/abs/1908.00986}
  {\bibinfo  {journal} {arXiv:1908.00986}\ }\BibitemShut {NoStop}%
\end{thebibliography}%

\newpage
\clearpage
\setcounter{figure}{0}
\setcounter{equation}{0}
\renewcommand{\theequation}{S\arabic{equation}}
\renewcommand{\thefigure}{S\arabic{figure}}
\renewcommand{\thesection}{S\arabic{section}}

\begin{widetext}
	
	\begin{center}
		\textbf{Supplemental Material}
	\end{center}

\section{Moir\'e Hamiltonian}

\begin{figure*}[b]
	\includegraphics[width=0.9\columnwidth]{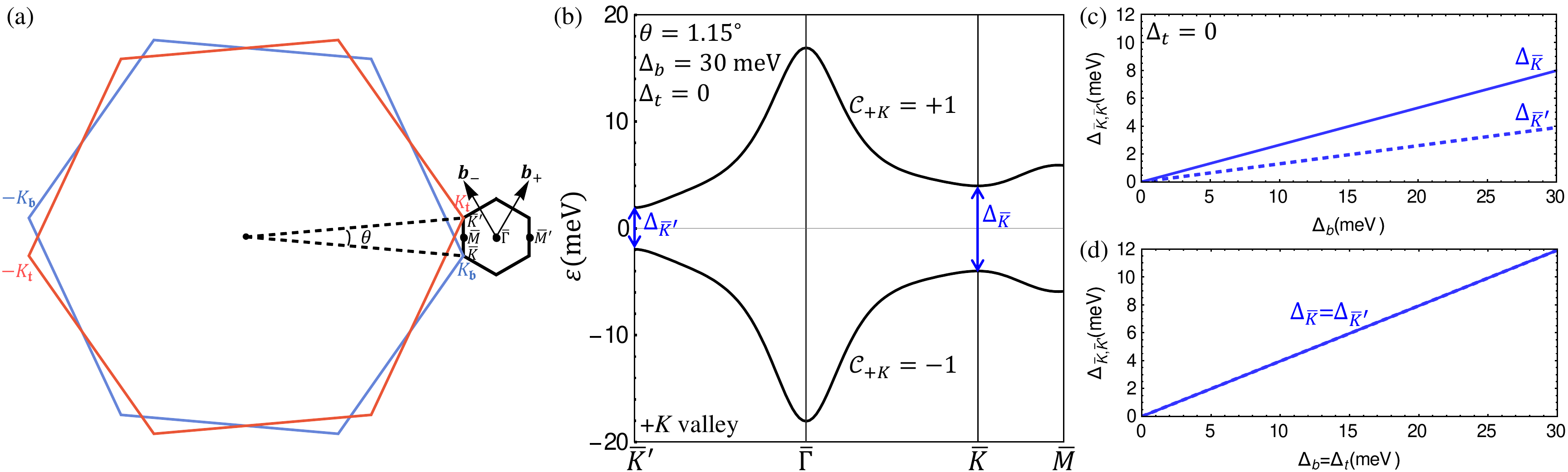}
	\caption{(a) Momentum space structure of twisted bilayer graphene. The blue and red hexagons represent the Brillouin zones associated respectively with the bottom and top layers. The black hexagon is the moir\'e Brillouin zone. (b) Moir\'e band structure in $+K$ valley with $\Delta_b = 30$ meV and $\Delta_t = 0$. This figure is the same as Fig.~1(a) in the main text. (c) $\Delta_{\bar{K}}$ and $\Delta_{\bar{K}'}$ as a function of $\Delta_b$, with $\Delta_t$=0. (d) Similar plot as (c) but with  $\Delta_b=\Delta_t$.  $\theta$ is set to $1.15^{\circ}$ in (b), (c) and (d).}
	\label{Fig:Gap0}
\end{figure*}

In twisted bilayer graphene (TBG) with a small twist angle $\theta$, the continuum moir\'e Hamiltonian \cite{Bistritzer2011} is given by
\begin{equation}
\mathcal{H}_{\tau}=\begin{pmatrix}
h_{\tau b}(\kk) & T_{\tau }(\rr) \\
T^{\dagger}_{\tau}(\rr) & h_{\tau t}(\kk)
\end{pmatrix},
\label{Hmoire}
\end{equation}
where $\rr$ and $\kk$ are respectively position and momentum operators, and  $\tau=\pm$ is the valley index.
$h_{\tau b}$ and $h_{\tau t}$ are the Dirac Hamiltonians of the bottom ($b$) and top ($t$) layers:
\begin{equation}
h_{\tau \ell}(\kk) = e^{-i\tau \ell \frac{\theta}{2} \sigma_z }[\hbar v_F (\kk-\tau \boldsymbol{\kappa}_{\ell})\cdot (\tau \sigma_x, \sigma_y)] + \Delta_{\ell} \sigma_z/2,
\label{hDirac}
\end{equation}
where $\ell$ is $+1$ ($-1$) for the $b$ ($t$) layer,
$v_F$ is the bare Dirac velocity($\sim 10^6$ m/s), and
$\sigma_{x, y, z}$ are Pauli matrices in the sublattice space.
Because of the relative rotation between the two layers, the Dirac cone in layer $\ell$ and valley $\tau$ is shifted to momentum
$\tau \boldsymbol{\kappa}_{\ell}=\tau [4\pi/(3 a_M)](-\sqrt{3}/2, -\ell/2)$, which is measured relative to the center of the moir\'e Brillouin zone [illustrated in Fig.~\ref{Fig:Gap0}(a)]. Here $a_M$ is the moir\'e period given by $a_0/[2 \sin (\theta/2)]$, where $a_0$ is the lattice constant of monolayer graphene. The term $\Delta_{\ell} \sigma_z/2$ in Eq.~(\ref{hDirac}) describes the sublattice potential difference in layer $\ell$. In pristine TBG, both $\Delta_{b}$ and $\Delta_{t}$ vanish. We assume that $\Delta_{b}$ ($\Delta_{t}$) can be induced when TBG is in close alignment to the bottom (top) hBN layers. Under this assumption, $\Delta_{b}$ and $\Delta_{t}$ can be independently controlled.

The interlayer tunneling terms $T_{\tau}(\rr)$ vary in space, following the periodicity of the moir\'e pattern:
\begin{equation}
T_{\tau}(\rr)=
T_{\tau}^{(0)}
+e^{-i \tau \bb_+ \cdot \rr} T_{\tau}^{(+1)}
+e^{-i \tau \bb_- \cdot \rr} T_{\tau}^{(-1)}
\end{equation}
where $\bb_{\pm}$ are moir\'e reciprocal lattice vectors given by $[4\pi/(\sqrt{3} a_M)](\pm1/2, \sqrt{3}/2)$ and
$T_{\tau}^{(j)} = w_{AA} \sigma_0 + w_{AB} \cos(2\pi j/3)\sigma_x+ \tau w_{AB}\sin(2\pi j/3) \sigma_y$. Here $w_{AA}$ and $w_{AB}$ are two parameters that respectively determine the tunneling in AA and AB/BA regions of the moir\'e pattern. We take $w_{AA} = 90$ meV and $w_{AB} = 117 $ meV \cite{wu2018topological, wu2019phonon}. 

The moir\'e Hamiltonian $\mathcal{H}_{\tau}$ is spin independent, and respects the spinless time-reversal symmetry that relates the two valleys. In the absence of  $\Delta_{b}$ and $\Delta_{t}$ , $\mathcal{H}_{\tau}$ builds in the $D_6$ point group symmetry that is generated by a sixfold rotation $\hat{C}_{6z}$ around the $\hat{z}$ axis and a twofold rotation $\hat{C}_{2x}$ around the $\hat{x}$ axis. Here the $\hat{C}_{2x}$ operation swaps the two layers.  For generic values of $(\Delta_{b}, \Delta_{t})$, the point group is reduced to $C_3$, since the twofold rotation $\hat{C}_{2z}$, which exchanges the two sublattices within each layer, is broken by the sublattice-dependent potentials $\Delta_{b,t}$. In the special case when $\Delta_{b}=\Delta_{t}$ ($\Delta_{b}=-\Delta_{t}$), the point group is $D_3$ that is generated by $\hat{C}_{3z}$ and $\hat{C}_{2y}$ ($\hat{C}_{2x}$).

The Dirac points located at the moir\'e Brillouin zone corners $\bar{K}$ and $\bar{K}'$ are gaped out when the $\hat{C}_{2z}$ symmetry is broken by $\Delta_{b,t}$. We denote the Dirac gap at $\bar{K}$ ($\bar{K}'$) in $+K$ valley  as $\Delta_{\bar{K}}$ ($\Delta_{\bar{K}'}$). If the interlayer tunnelings in TBG were absent, we would find that $\Delta_{\bar{K}}=\Delta_b$ and $\Delta_{\bar{K}'}=\Delta_t$. Due to the interlayer tunnelings, wave functions of TBG are different from those of two decoupled monolayer graphene, and therefore, $\Delta_{\bar{K}}$ ($\Delta_{\bar{K}'}$) deviates from $\Delta_b$ ($\Delta_t$). In Fig.~\ref{Fig:Gap0}(c), we plot $\Delta_{\bar{K}}$ and $\Delta_{\bar{K}'}$ as a function of $\Delta_b$, while $\Delta_t$ is set to 0. In this case $\Delta_{\bar{K}}$ and $\Delta_{\bar{K}'}$ take different values, and both are finite, but smaller compared to $\Delta_b$. Fig.~\ref{Fig:Gap0}(d) is a similar plot but with $\Delta_b=\Delta_t$, and therefore, $\Delta_{\bar{K}} = \Delta_{\bar{K}'}$.

We obtain the wave function $\Phi_{\kk, \tau}$ of the moir\'e band by diagonalizing the Hamiltonian $\mathcal{H}_{\tau}$ using a plane-wave expansion. The wave function $\Phi_{\kk, \tau}$ contains the symmetry properties discussed above.

\begin{figure*}[t]
	\includegraphics[width=0.9\columnwidth]{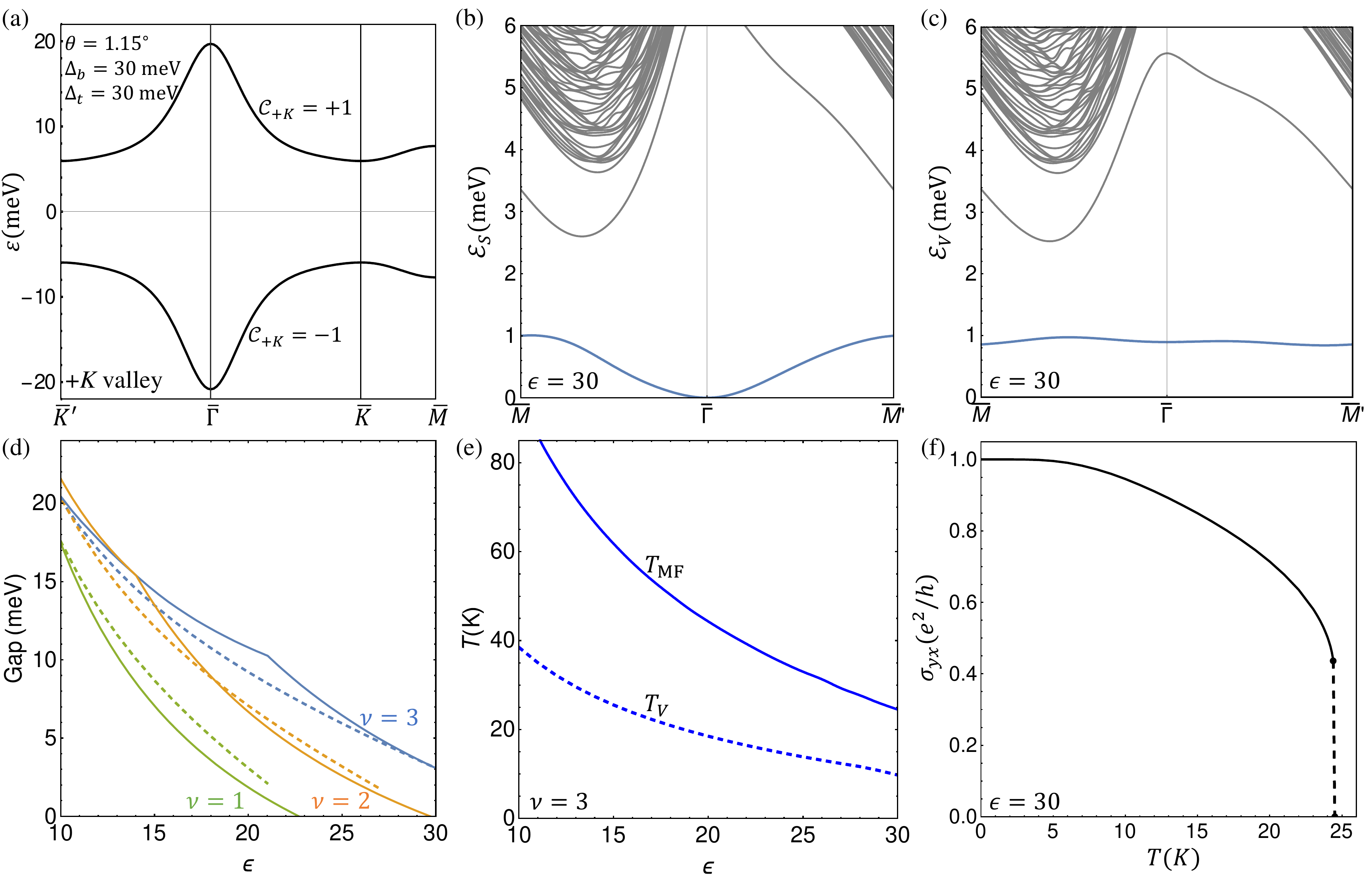}
	\caption{Results by taking $\Delta_b=\Delta_t=30$ meV. (a) Moir\'e band structure. (b) Spin magnon spectrum at $\nu=3$. (c) Valley magnon spectrum at $\nu=3$. (d) Charged excitation gap as a function of dielectric constant $\epsilon$. The solid lines represent $\Delta_{\text{HF}}$ respectively for the three integer filling factors, and the dashed lines the skyrmion-antiskyrmion pair energy $\Delta_{\text{pair}}$ assuming spin maximally polarized state at each filling factor. This plot shows that $\Delta_{\text{pair}}$ can be smaller than the corresponding $\Delta_{\text{HF}}$. (e) Transition temperature at $\nu = 3$ as a function of $\epsilon$. The solid line shows the mean-field transition temperature $T_{\text{MF}}$, and the dashed line the valley ordering temperature $T_V$ estimated using the valley wave spectrum. (f) Mean-field value of the anomalous Hall conductivity $\sigma_{yx}$ at $\nu=3$ as a function of temperature.}
	\label{Fig:banddouble}
\end{figure*}

\section{Quantum Anomalous Hall Ferromagnets for $\Delta_{b}=\Delta_{t}$}
In our study of interaction effects, we only keep the first moir\'e conduction band states. This approximation becomes better when the Coulomb interaction strength becomes much smaller than the band gaps to remote moir\'e bands. The characteristic interaction strength is set by $E_C= e^2/(\epsilon a_M)$, where $\epsilon$ is the dielectric constant that can be controlled by the three-dimensional dielectric environment. Using $\epsilon=30$ and $\theta=1.15^{\circ}$, we find that $a_M \approx 12.3$ nm and $E_C \approx 3.9$ meV. In our moir\'e band structure, the band gap that separates the first and the second conduction bands is about 40 meV, which is an order of magnitude larger than the above interaction scale. The band gap that separates the first conduction band and the first valence band is the charge neutrality gap $\Delta_{\nu=0}$, which is approximately equal to $\min \{\Delta_{\bar{K}}, \Delta_{\bar{K}'}\}$. In the main text, we take $(\Delta_{b}, \Delta_{t})= (30, 0)$ meV, and $\Delta_{\nu=0}$ is then about 4 meV, which is, however, only slightly larger than $E_C$ estimated above. We note the corresponding experimental value of  $\Delta_{\nu=0}$ is about 6 meV \cite{serlin2019intrinsic}.

Here we propose that $\Delta_{\nu=0}$ can be enhanced when both top and bottom hBN layers have a zero orientation angle relative to TBG.  As shown in Figs.~\ref{Fig:Gap0}(d) and \ref{Fig:banddouble}(a), $\Delta_{\nu=0}$ is enhanced to 12 meV by taking $\Delta_{b} = \Delta_{t} =30$ meV, and becomes three times the interaction scale $E_C(\epsilon =30)$, which makes the projection of interaction to the first conduction band a better approximation. For this TBG system with an increased $\Delta_{\nu=0}$, we perform the same analysis on interaction effects as in the main text, with results summarized in Fig.~\ref{Fig:banddouble}. We find that the main conclusions remain unchanged: (1) The quantum anomalous Hall ferromagnets (QAHF) at $\nu=3$ is generally robust against spin wave and valley wave excitations, provided that the Hartree-Fock gap is finite, as shown in Figs.~\ref{Fig:banddouble}(b) and \ref{Fig:banddouble}(c); (2) The Curie temperature is still limited by valley wave excitations, and is reduced from the mean-field value [Fig.~\ref{Fig:banddouble}(e)]. However, there is an important difference regarding the skyrmion-antiskyrmion pair energy $\Delta_{\text{pair}}$. In particular, we find that $\Delta_{\text{pair}}$ can be lower than $\Delta_{\text{HF}}$ for certain values of $\epsilon$, as shown in Fig.~\ref{Fig:banddouble}(d). This is partially because the first moir\'e conduction band becomes narrower by taking $\Delta_{b} = \Delta_{t} =30$ meV. Therefore, the skyrmion-antiskyrmion pairs can be the lowest charged excitation, depending on the details of the model. 

The results shown in Fig.~\ref{Fig:banddouble} provide further theoretical evidences that the TBG QAHF are robust against particle-hole fluctuations. When the parameter $\Delta_{\nu=0}/E_C$ decreases, we expect that the QAHF should remain robust within a finite range of $\Delta_{\nu=0}/E_C$, even after effects of other bands are taken into account. An interesting question is whether there exists a critical value of $\Delta_{\nu=0}/E_C$, below which the QAHF becomes unstable. This question is beyond the scope of the current work, and we leave it to future study.

In Fig.~\ref{Fig:banddouble}(d), the Hartree-Fock gap $\Delta_{\text{HF}}$ at $\nu=3$ as a function of the dielectric constant $\epsilon$ has a kink around $\epsilon \approx 21$. There are similar kinks in Fig. 2(a) of the main text. The reason for this kink can be explained as follows. $\Delta_{\text{HF}}$ at $\nu=3$ is defined as $E_{+,\downarrow}^{(\text{min})} - E_{+,\uparrow}^{(\text{max})}$, where $E_{+,\downarrow}^{(\text{min})}$ is the minimum quasiparticle energy of the unoccupied spin $\downarrow$ band at valley $+K$, and $E_{+,\uparrow}^{(\text{max})}$ is the maximum quasiparticle energy of the occupied band. At small $\epsilon$ (i.e., large Coulomb interaction), $E_{+,\downarrow}^{(\text{min})}$ occurs at the center of MBZ due to a strong modification of the quasiparticle energy bands by the Coulomb interaction. By contrast, at large $\epsilon$ (i.e., small Coulomb interaction), $E_{+,\downarrow}^{(\text{min})}$ appears at the corners of MBZ. Therefore,  $\Delta_{\text{HF}}$  as a function of  $\epsilon$ can have a kink.

\end{widetext}

\end{document}